\documentclass[a4paper,11pt]{article}
\usepackage{jheppub} 
\pdfoutput=1
\usepackage[utf8]{inputenc}
\usepackage{hyperref}
\usepackage{amsfonts}
\usepackage{amsmath}
\usepackage{amssymb}
\usepackage{mathrsfs}  
\usepackage{color}
\usepackage{physics}
\usepackage{multicol}
\usepackage{tikz}

\usepackage{amsmath,bbm,array,amsfonts,graphicx,wrapfig,lscape,float,mathtools,multirow,longtable,amsthm}
\usepackage{stackrel}
\usepackage[all]{xy}
\usepackage{caption}
\usepackage{subcaption}
\usepackage{multirow}
\usepackage[bottom]{footmisc}
\usepackage{mathrsfs}
\usepackage{tikz}
\usepackage{bm}
\usepackage{color}
\usepackage{enumerate}
\usetikzlibrary{decorations.pathreplacing}
\usepackage{diagbox}
\numberwithin{equation}{section}
\numberwithin{figure}{section}
\numberwithin{table}{section}
\usepackage{pgfplots}
\pgfplotsset{compat=1.14}
\usepackage{makecell}
\usepackage{lscape}
\usepackage[utf8]{inputenc}
\usepackage{mathtools}
\usepackage{todonotes}
\usepackage{epigraph}
\usepackage{boldline}

\usetikzlibrary{decorations.pathmorphing}
\usetikzlibrary{decorations.pathreplacing}
\usetikzlibrary{decorations.markings}
\usetikzlibrary{shapes, shapes.geometric, shapes.symbols, shapes.arrows, shapes.multipart, shapes.callouts, shapes.misc}
\usetikzlibrary{shadows,fit,trees,calc,patterns,arrows}
\tikzset{snake it/.style={decorate, decoration=snake, segment length=2mm}}
\tikzset{7brane/.style={circle, draw=black, fill=black,ultra thick,inner sep=1.5 pt, minimum size=1 pt,}, c/.default={4pt}}
\tikzset{cross/.style={cross out, draw=black,ultra thick, minimum size=2*(#1-\pgflinewidth), inner sep=0pt, outer sep=0pt}, cross/.default={5pt}}

\newtheorem{definition}{Definition}[section]
\newtheorem{theorem}{Theorem}[section]

\newcommand{\CC}{\mathbb{C}}
\newcommand{\ZZ}{\mathbb{Z}}
\newcommand{\QQ}{\mathbb{Q}}

\begin{document}

\title{Machine Learning Algebraic Geometry for Physics}

\author[a,b]{Jiakang Bao}
\author[b,a,c,d]{Yang-Hui He}
\author[a,b]{Elli Heyes}
\author[a,b]{Edward Hirst}

\affiliation[a]{Department of Mathematics, City, University of London, EC1V 0HB, UK}
\affiliation[b]{London Institute for Mathematical Sciences, Royal Institution, London W1S 4BS, UK}
\affiliation[c]{Merton College, University of Oxford, OX1 4JD, UK}
\affiliation[d]{School of Physics, NanKai University, Tianjin, 300071, P.R. China}

\emailAdd{jiakang.bao@city.ac.uk}
\emailAdd{hey@maths.ox.ac.uk}
\emailAdd{elli.heyes@city.ac.uk}
\emailAdd{edward.hirst@city.ac.uk}

\preprint{LIMS-2022-012}

\abstract{We review some recent applications of machine learning to algebraic geometry and physics. Since problems in algebraic geometry can typically be reformulated as mappings between tensors, this makes them particularly amenable to supervised learning. Additionally, unsupervised methods can provide insight into the structure of such geometrical data.
At the heart of this programme is the question of how geometry can be machine learned, and indeed how AI helps one to do mathematics.

This is a chapter contribution to the book {\it Machine learning and Algebraic Geometry}, edited by A.~Kasprzyk et al. \\

\emph{To the memory of Professor John K. S. McKay (1939-2022), with deepest respect.}
}

\maketitle

\section{Introduction}\label{sec:intro}
The ubiquitous interrelations between algebraic geometry and physics has for centuries flourished fruitful phenomena in both fields.
With connections made as far back as Archimedes whose work on conic sections aided development of concepts surrounding the motion under gravity, physical understanding has largely relied upon the mathematical tools available.
In the modern era, these two fields are still heavily intertwined, with particular relevance in addressing one of the most significant problems of our time - quantising gravity.
String theory as a candidate for this theory of everything, relies heavily on algebraic geometry constructions to define its spacetime and to interpret its matter.

However, where new mathematical tools arise their implementation is not always simple. 
Current techniques for large scale statistical analysis would be at a loss without the recent exponential growth in the power of computation.
It is hence only natural to investigate the prosperity of big data techniques in guiding this innate partnership between geometry and physics; and for this we focus on a big data key player: machine learning (ML).

Machine learning has seen a vast array of applications to algebraic geometry in physics, with first introduction to the string landscape in \cite{He:2017aed,Carifio:2017bov,Krefl:2017yox,Ruehle:2017mzq,He:2017set}, and numerous current uses in finding suitable metrics on the landscape \cite{Douglas:2006rr,Bull:2018uow,Bull:2019cij,He:2020lbz,Anderson:2020hux,Jejjala:2020wcc,Douglas:2020hpv,Erbin:2021hmx,Larfors:2021pbb,He:2021eiu,Abel:2021rrj,Constantin:2021for,ashmore2021machine,gao2021machine,Cole:2021nnt,He:2021eiu}.
More generally, the question of machine-learning mathematics and using data scientific methods as well as artificial intelligence to help humans to do mathematics is very much in the air \cite{He:2018jtw,He:2019nzx,He:2020fdg,buzzard2020proving,He:2021oav,davies2021advancing,He:2022cpz}.

Many of these methods use neural networks (NN) in some form as generalisations of the multi-layer perceptron (MLP) for non-linear function approximation.
In any respect, new approaches to mathematical physics will continue to rely further on techniques that can suitably handle large amounts of data.
Pattern recognition for conjecture formulation may even eventually become more efficient with ML techniques through clustering and degree of freedom identification with methods such as principal component analysis (PCA).

In this work we review a selection of contemporary studies into the use of ML in algebraic geometry for physics.
Specifically each section finds focus on application to: \S\ref{mlhypersurfaces} hypersurfaces \cite{Berman:2021mcw}, \S\ref{polytopes} polytopes \cite{Bao:2021ofk}, \S\ref{HS} Hilbert series \cite{Bao:2021auj}, \S\ref{amoeba} amoebae \cite{Bao:2021olg}, \S\ref{branewebs} brane webs \cite{Arias-Tamargo:2022qgb}, \S\ref{quivermutation} quiver mutation \cite{Bao:2020nbi}, \S\ref{dessins} dessins d'enfants \cite{He:2020eva}, and \S\ref{hessiangan} Hessian manifolds.

\paragraph{The string landscape} Superstring theory requires the spacetime to be of (real) dimension 10. To reconcile with our experience of 4-dimensional spacetime, the extra dimensions must be curled up so that they cannot be observed at low energy. In other words, our theory is compactified on a 6-dimensional manifold $\mathcal{M}_6$ where the full spacetime is $\mathcal{M}_{10}=\mathbb{R}^{3,1}\times\mathcal{M}_6$. In fact, $\mathcal{M}_6$ can be endowed with complex structure, and we shall henceforth refer to it as a threefold in terms of its complex dimension. As pointed out in \cite{Candelas:1985en}, a standard solution for this threefold is a Ricci flat K\"ahler manifold. Such manifolds are famously known as the Calabi-Yau (CY) manifolds.

The physics in the Minkowski spacetime $\mathbb{R}^{3,1}$ is dictated by the geometry and topology of the CY threefolds. It is still not clear which CY$_3$ is \emph{the} manifold for string theory compactification. Nevertheless, people have made great efforts in finding distinct CY threefolds over the past few decades. Nowadays, the full list contains over billions of possible candidates. In the era of big data, the technique of machine learning therefore naturally enters this string landscape. For recent summaries and reviews, see \cite{He:2018jtw,Bao:2020sqg,He:2020bfv,He:2020mgx,He:2021oav,He:2022cpz}.

We should emphasize that the concept of CY $n$-fold in this note is defined in a rather broad sense. We shall not only discuss compact CYs, but also take non-compact (also known as ``local'') ones into account. Physically, one can turn on the $\Omega$-background which localizes our theory on the non-compact CY and effectively compactifies it to 4d. Moreover, we will also consider singular algebraic varieties, especially Gorenstein singularities which can be resolved to smooth CYs.

\paragraph{Machine Learning} 
Most machine learning algorithms can be put into two categories: unsupervised learning and supervised learning methods (there are other categories such as reinforcement learning and semi-supervised learning). Unsupervised methods are used when one is given unlabelled data $\{x_{i}\}$ and wishes to find patterns/structures within the data. These methods can be broken down further into clustering methods, such as $k$-means clustering and dimensionality reduction methods, such as principal component analysis (PCA). Supervised learning on the other hand, is used when one has labeled data $\{(x_{i},y_{i})\}$ and wishes to approximate a mapping function $f:x_{i} \mapsto y_{i}$, such methods include linear regression, neural networks (NN), of which convolutional neural networks are a subtype (CNN), random forests, and support vector machines. Supervised learning can also be broken down further into regression and classification problems. In Regression problems the output variables $y$ are continuous numerical values whereas in classification problems $y$ is discrete and not necessarily numerical. For regression problems common performance measures include mean absolute error (MAE) and mean squared error (MSE), for classification problems accuracy is often used and Matthew's Correlation Coefficient (MCC) is another popular choice where the classes are of different sizes.

In order to get an unbiased evaluation of how well your supervised learning algorithm is learning to map the data, the data is shuffled and then separated into training and test sets. The training dataset is used to fit the model parameters, and the test dataset is used to provide an unbiased evaluation of the model. The usual train:test split is 80:20. Cross-validation is another method used to obtain an unbiased evaluation of the model performance, whereby the data is first shuffled and split into $k$ groups, then each group is taken in turn to be the test set and the remaining groups are combined to create the training set. In each case the model is trained on the training set, evaluated on the test set, and the evaluation score is recorded. The mean and standard deviation of the $k$ evaluation scores are then used to determine the model performance. 

For the projects reviewed here, most work was carried out in \texttt{Python} with the use of \texttt{TensorFlow} \cite{tensorflow2015-whitepaper} and \texttt{scikit-learn} \cite{scikit-learn} for machine learning implementation, \texttt{Yellowbrick} \cite{bengfort_yellowbrick_2018} for visualisations and \texttt{ripser} \cite{ctralie2018ripser} for topological data analysis.


\section{Calabi-Yau Hypersurfaces}\label{mlhypersurfaces}
There have been many very interesting applications of machine learning to the Calabi-Yau landscape. Here we focus on summarising the work in \cite{Berman:2021mcw} which centres its attention on a specific subset of the landscape: Calabi-Yau 3-folds constructed from hypersurfaces in weighted projective space, $\mathbb{P}^4$.

One of the first big datasets in theoretical physics came from a generalisation of the quintic 3-fold in $\mathbb{P}^4$. 
The generalisation assigned weights to the respective identification used in defining the projective space, such that for some $\mathbb{C}^{n+1}$ with coordinates $\{z_1,z_2,...,z_{n+1}\}$, a set of $n+1$ coprime integer weights, $w_i$, could be selected to describe the identification
\begin{equation}\label{eq:weightedprojective}
    (z_1,z_2,...,z_{n+1}) \sim (\lambda^{w_1}z_1,\lambda^{w_2}z_2,...,\lambda^{w_{n+1}}z_{n+1})\,,
\end{equation}
$\forall \lambda \in \mathbb{C}$. These, now weighted, projective spaces can have codimension-1 hypersurfaces drawn within them, and it turns out for only a specific set of 7555 weight combinations, for the $n=4$ case, can one define hypersurfaces within them which are Calabi-Yau in nature \cite{CANDELAS1990383}.
This construction method for Calabi-Yau 3-folds now forms a distinguished subset of the full Kreuzer-Skarke database \cite{Kreuzer:2000xy}, which describes Calabi-Yau hypersurfaces in more general toric varieties.

Why only these specific 7555 weight combinations permit Calabi-Yau hypersurfaces is perplexing. 
Beyond the defining Calabi-Yau feature of vanishing first Chern class causing the hypersurface polynomial to have degree $\sum_iw_i$, there is a necessary but not sufficient condition on the weights for the hypersurfaces to be consistently defined over the singular sets created by the projective identification. 
This condition is \textit{transversity} and sets a specific divisibility condition on the weights.

To examine this database the work of \cite{Berman:2021mcw} turned to tools from supervised and unsupervised ML, which we summarise in this section. 
Unsupervised methods sought to provide general data analysis of this special database; whilst supervised methods aimed to learn the topological properties of the hypersurfaces directly from the weights. 
The learning hence emulating the highly non-trivial formulas from the Poincaré polynomial, $Q(u,v)$, expansion which provides the Hodge numbers, $h^{\alpha,\beta}$, and a direct equation for the Euler number, $\chi$.
\begin{equation}\label{eq:hodgeeulerformulas}
\begin{split}
    Q(u,v)&= \sum_{\alpha,\beta} h^{\alpha,\beta}u^\alpha v^\beta= \frac{1}{uv} \sum_{l=0}^{\sum\limits_i(w_i)} \bigg[ \prod_{\tilde{\theta}_i(l)\in\mathbb{Z}} \frac{(uv)^{q_i}-uv}{1-(uv)^{q_i}} \bigg]_\text{int} \bigg( v^{\text{size}(l)} \bigg(\frac{u}{v}\bigg)^{\text{age}(l)}\bigg) \;,\\
    \chi&= \frac{1}{\sum\limits_i(w_i)} \sum_{l,r=0}^{\sum\limits_i(w_i)-1} \bigg[\prod_{i|lq_i , rq_i \in \mathbb{Z}} \bigg(1-\frac{1}{q_i}\bigg)\bigg]\;,
\end{split}
\end{equation}
such that normalised weights $q_i = w_i/\sum\limits_i(w_i)$ are operated on with multiple divisibility checks and other complicated computation steps involving age$(l) = \sum_{i=0}^4 \tilde{\theta}_i(l)$ and size$(l) = \text{age}(l) + \text{age}(\sum_i(w_i) - l);$ where in $(\mathbb{R}/\mathbb{Z})^5, \ \tilde{\theta}_i(l)$ is the canonical representative of $lq_i$  \cite{Vafa:1989xc,Batyrev:2020ych}.

\subsection{PCA, Clustering, and TDA}
The dataset of weighted-$\mathbb{P}^4$'s which admit Calabi-Yau hypersurfaces takes the form of 7555 sorted 5-vectors of positive integer weights.
In addition to this information, the non-trivial Hodge numbers $\{h^{1,1},h^{2,1}\}$ of each hypersurface are provided with these weights of the ambient projective variety in the Kreuzer-Skarke database: \url{http://hep.itp.tuwien.ac.at/~kreuzer/CY/}. 

For dataset feature comparison, and to benchmark learning performance, three further datasets were generated of: 1) Random positive integers, 2) Coprime positive integers, 3) Transverse coprime integers. 
Therefore with the Calabi-Yau weight data these datasets form a series, each with more of the necessary (but not sufficient) conditions for the `Calabi-Yau property'. 
Furthermore these datasets were constructed so as to have no overlap with each of the other datasets in the series, therefore not satisfying the properties at the next level and hence ensuring artefacts of the analysis could be attributed to the property inclusion.

\paragraph{Principal Component Analysis}
To determine structure in the weight vector data, PCA was applied to each of the datasets. 
This method finds the linear combinations of the weights which dictate the most dominant variation in the data.
The process diagonalises the covariance matrix such that eigenvalues give successively lower variances and the respective eigenvectors are the principal components which best describe the distribution of the weights.

\begin{figure}[h!]
	\centering
	\begin{subfigure}{0.45\textwidth}
    	\centering
    	\includegraphics[width=0.65\textwidth]{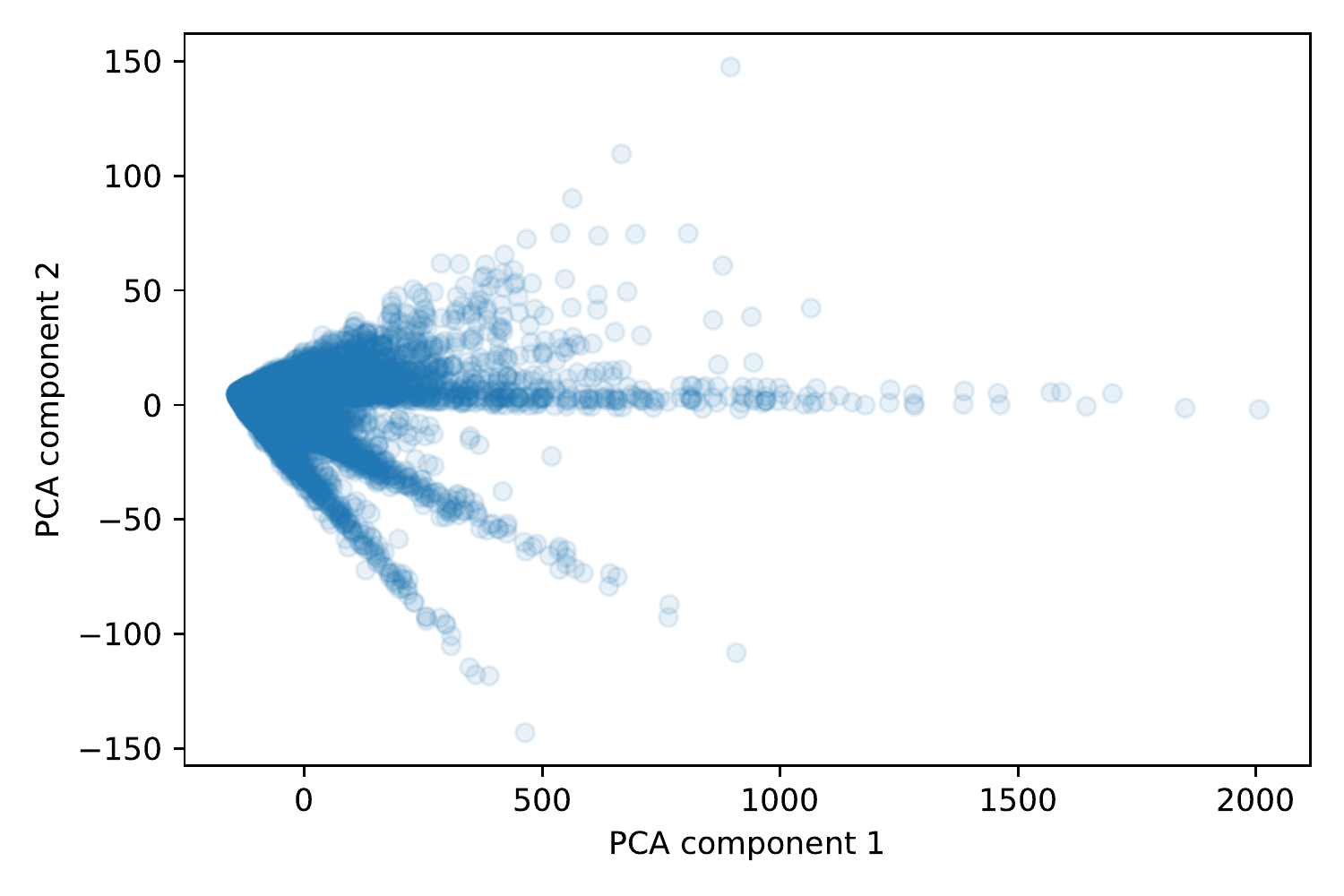}
    	\caption{2d PCA}\label{fig:CYPCA}
	\end{subfigure} 
    \begin{subfigure}{0.45\textwidth}
    	\centering
    	\includegraphics[width=0.65\textwidth]{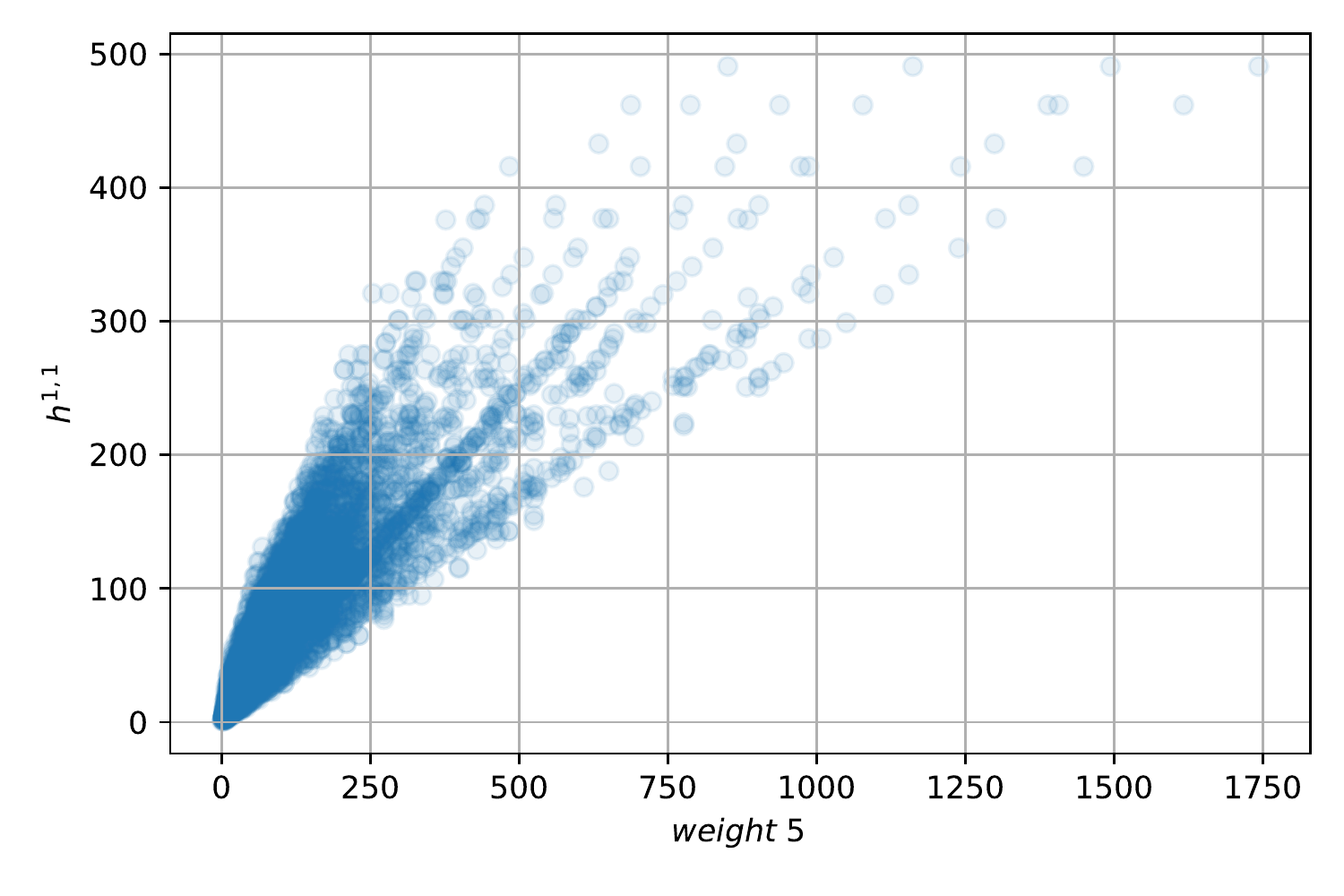}
    	\caption{$w_5$ vs $h^{1,1}$}\label{fig:w5_h11}
    \end{subfigure}
\caption{Linear clustering behaviour shown from both the 2d PCA projection of the dataset (a), and the distribution of largest ambient space weight against $h^{1,1}$ of the hypersurface (b).}\label{fig:CYPCA_h1w5}
\end{figure}

Since the weight vectors are sorted in increasing order, the principal components have a vague correlation with the weights themselves, however as the PCA plot in Figure \ref{fig:CYPCA} shows the Calabi-Yau data has a clear linear clustering structure in the first two principal components, not present in the other generated datasets.

\paragraph{Clustering}
Plotting the weights of each ambient space against the Hodge numbers of the Calabi-Yau hypersurface exhibits a familiar behaviour to that seen in the PCA, as shown in \ref{fig:w5_h11}.
The linear clustering behaviour observable by eye suggests a surprising simpler structure in the data which ML methods can make use of for approximate learning.
However prior to examining the supervised learning in \ref{sec:NN_hypersurfaces}, the clustering behaviour can be quantitatively verified with the use of the unsupervised method K-Means clustering.

K-Means clustering takes as input a number to define how many clusters to compute, then iteratively updates the cluster centres until the average distance between any datapoint and its nearest cluster centre is minimised. 
The number of clusters was determined using a variant of traditional elbow methods, recommending an optimum of 10 clusters for the clustering algorithm.
This distance measure is known as \textit{inertia}, and an equivalent version normalised by the number of clusters and datapoints gives a final metric to establish how suitable this clustering behaviour was.

For this 2d $w_5$ vs $h^{1,1}$ data, to evaluate the linear behaviour each datapoint was projected onto its ratio $h^{1,1}/w_5$, and then clustering performed on this now 1d data. 
The final normalised inertia took value: 0.00084; which may be interpreted as after determining optimal cluster centres each datapoint is on average less than 0.1\% of the full data range from its nearest cluster centre. 
These are exceptionally good clustering results and strongly corroborate the behaviour observed.

\paragraph{Topological Data Analysis}
Persistent homology is a tool form topological data analysis most useful for analysis of higher-dimensional data structures that cannot be plotted directly.
Since the weight data here is 5 dimensional it is perfectly suited for this line of analysis also. For other uses of TDA in the theoretical physics please see \cite{Cirafici:2015pky,Cole:2018emh}.

The process involves the plotting of the 5d data in $\mathbb{R}^5$, and drawing 5d balls centred on each point all with radius $d$. 
As the ball radii are varied from $0 \mapsto \infty$ they begin to intersect, a $n$-simplex is then drawn between $n$ points when their balls intersect; note in 5d only up to 5-simplices can be drawn. 
This forms a chain of Vietoris-Rips complexes for each $d$ value which changes the number of intersections.

The homology of this chain can then be computed, with the use of the \texttt{python} library \texttt{ripser} \cite{ctralie2018ripser}; and due to computational memory restrictions we limit analysis to $H_0$ and $H_1$ in this study.
$H_0$ features are each of the points individually, hence all `born' at $d=0$, and die as they become connected components (the usual convention for the feature allocated to the smaller component dying as two become connected).
These features hence provide useful information on the distribution of datapoints.
Conversely $H_1$ features are sets of 1-simplices (edges) that form the boundary of a union of 2-simplices not in the complex.
They keep track of 2d holes in the data and gives further useful information about the datapoint distribution.

\begin{figure}
    \centering
    \includegraphics[width=0.4\textwidth]{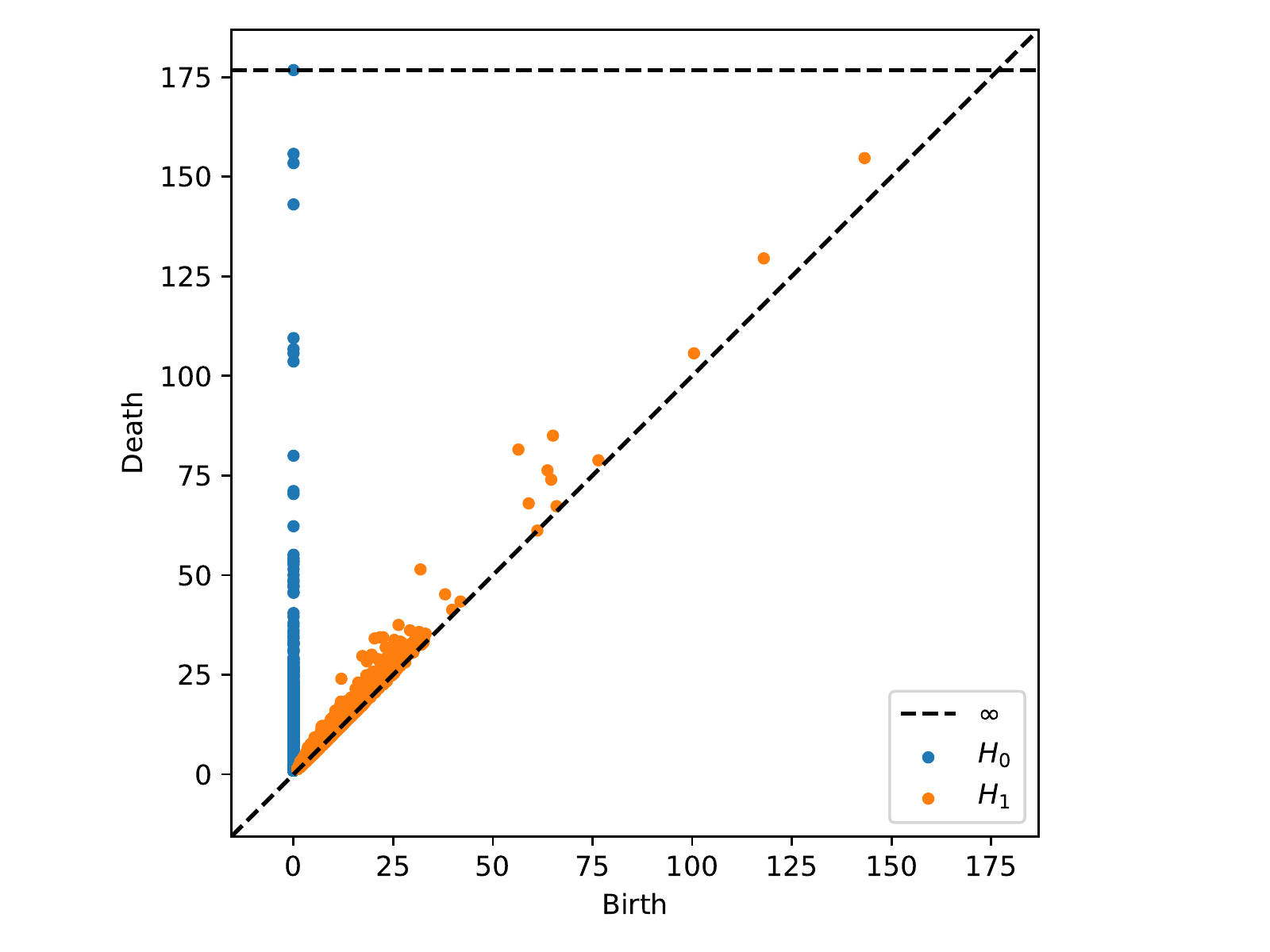}
    \caption{Persistent diagram for the $H_0$ and $H_1$ homology groups of the CY data's Vietoris-Rips complex filtration.}
    \label{fig:CYTDA}
\end{figure}

The $H_i$ features are then plotted as (birth,death) pairs on a persistence diagram, for the Calabi-Yau weight data this is shown in Figure \ref{fig:CYTDA}.
The $H_0$ features show gaps in the line, indicating that some datapoints are separated in some way into clusters which join together at higher $d$ values.
This behaviour is expected since some datapoints have particularly high $w_5$ values, causing them to be significantly further away from the bulk of the data, likely connecting amongst themselves along the linear clusters before joining the bulk complex.
The $H_1$ features all lie close to the diagonal, behaviour typical of noise, and thus there is not any further higher-dimensional structure in the data connecting these classes together to form persistent loops.
Both these result support the observed clustering behaviour.

\subsection{Neural Networks for Hypersurfaces}\label{sec:NN_hypersurfaces}
For the supervised learning of this data, two investigations were carried out, one using regressors to predict topological parameters of the hypersurface, and the other classifiers to distinguish weight vectors that admit Calabi-Yau hypersurfaces from those which do not.

\paragraph{Regression}
The first used a typical feed-forward neural network regressor to learn the Hodge numbers and Euler number from the 5 vectors of weights, using the standard mean squared error loss.
For Calabi-Yau 3-folds there is a simple expression for the Euler number in terms of the Hodge numbers: $\chi = 2(h^{1,1}-h^{2,1})$.
Performance was best measured on the independent test data using the $R^2$ metric, determining the performance of the network with respect to a null model always predicting the mean output. 
The metric evaluates in the range $(-\infty,1]$ with value 1 for perfect prediction.
To provide confidence in the measure 5-fold cross-validation was also performed such that 5 identical architectures were independently trained and tested on 5 different partitions of the data. 

The results for this investigation are given in Table \ref{tab:ML_weights}. They show very strong learning with $R^2$ values close to 1. Since the formulas \ref{eq:hodgeeulerformulas} require a lot of divisibility checks which NNs are notoriously poor at, it is likely the networks are making use of some other structure, perhaps the linear clustering observed, to predict these topological parameters.

\begin{table}[h!]
\centering
\begin{tabular}{|c|cccc|}
\hline
\multirow{2}{*}{Measure} & \multicolumn{4}{c|}{Property} \\ \cline{2-5} & \multicolumn{1}{c|}{$h^{1,1}$} & \multicolumn{1}{c|}{$h^{2,1}$} & \multicolumn{1}{c|}{$[h^{1,1},h^{2,1}]$} & $\chi$ \\ \hline
$R^2$                       & \multicolumn{1}{c|}{\begin{tabular}[c]{@{}c@{}}0.9630\\ $\pm$ 0.0015\end{tabular}} & \multicolumn{1}{c|}{\begin{tabular}[c]{@{}c@{}}0.9450\\ $\pm$ 0.0133\end{tabular}} & \multicolumn{1}{c|}{\begin{tabular}[c]{@{}c@{}}0.9470\\ $\pm$ 0.0041\end{tabular}} & \begin{tabular}[c]{@{}c@{}}0.9510\\ $\pm$ 0.0023\end{tabular} \\ \hline
\end{tabular}
\caption{Learning the non-trivial Hodge numbers and Euler number from the Calabi-Yau 5-vectors of weights. Measurement of learning performance uses 5-fold cross-validation to provide an average and standard error on each measure's value.}
\label{tab:ML_weights}
\end{table}

\paragraph{Classification}
The second investigation used a variety of classification architectures (including NNs), most notably performance was not significantly improved on by the more complicated architectures compared to the simple prototypical classification architecture of Logistic Regression.
Distinguishing each of the generated datasets of weights from the Calabi-Yau dataset of weights gave accuracies $\sim 0.7$ in all cases, showing some learning, particularly impressive for the case where both datasets exhibited transversity. 

\begin{figure}[h!]
	\centering
	\begin{subfigure}{0.45\textwidth}
    	\centering
    	\includegraphics[width=0.8\textwidth]{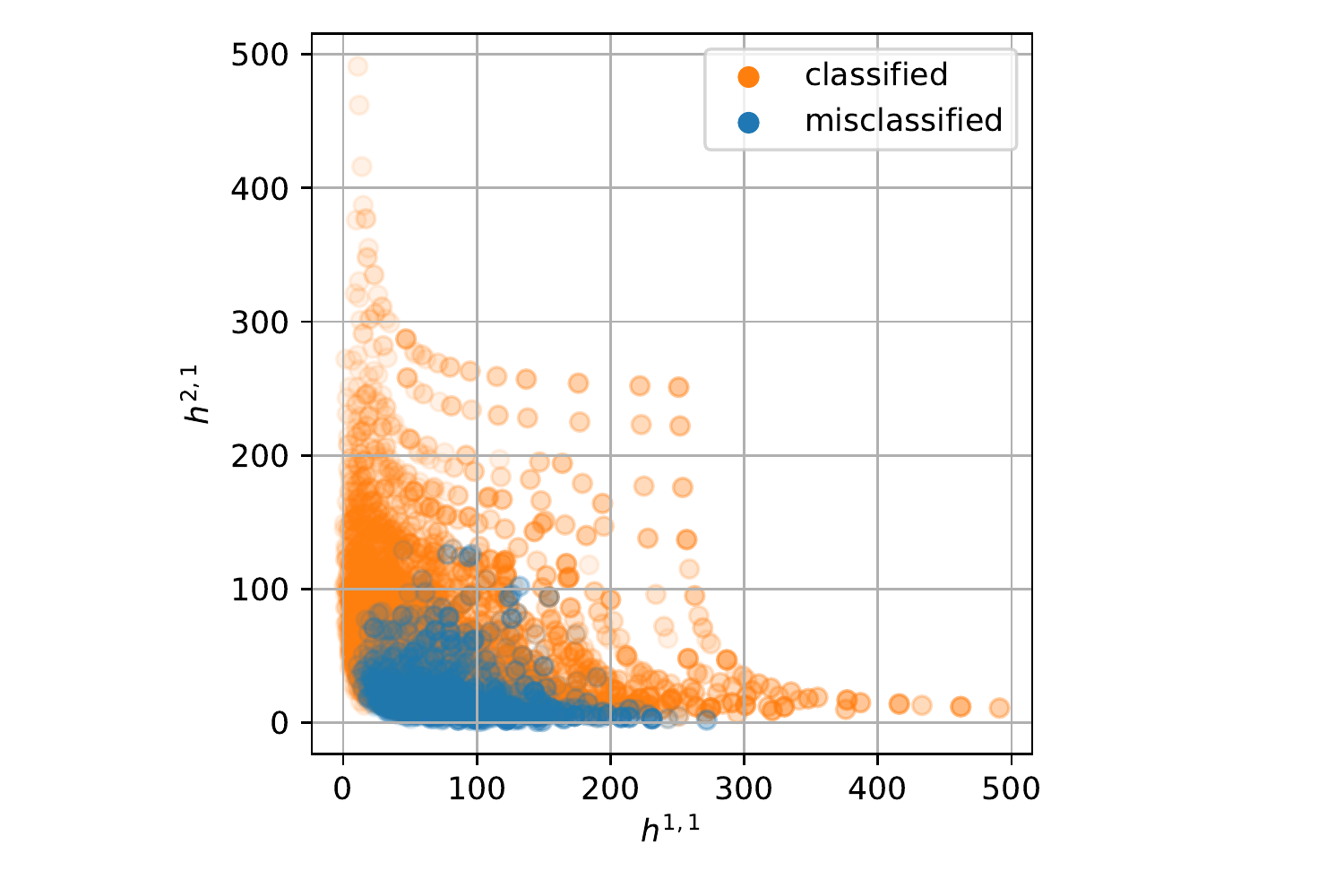}
    	\caption{Train with Random}\label{fig:LRR_misclass}
	\end{subfigure} 
    \begin{subfigure}{0.45\textwidth}
    	\centering
    	\includegraphics[width=0.8\textwidth]{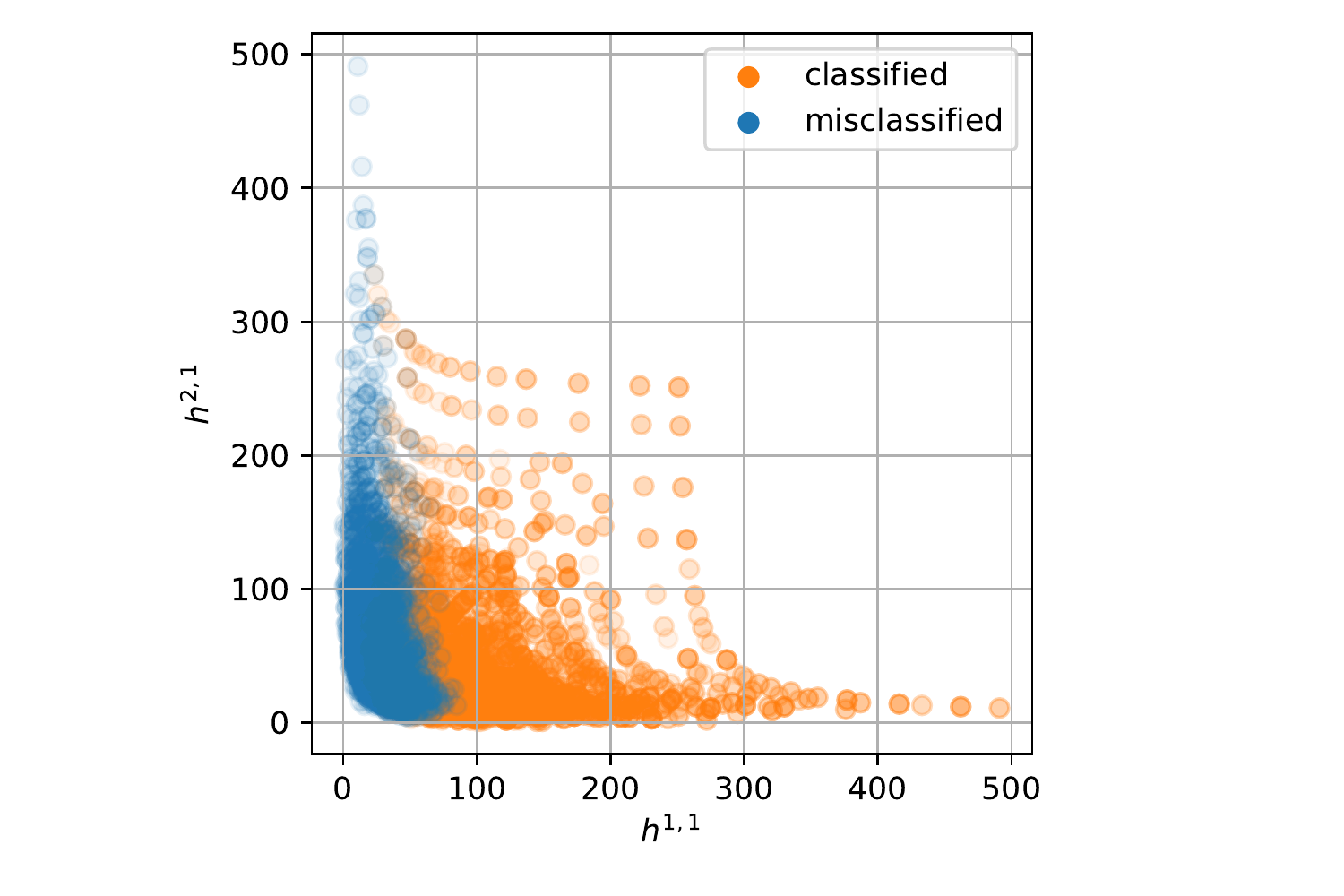}
    	\caption{Train with Transverse}\label{fig:LRT_misclass}
    \end{subfigure}
\caption{The Logistic Regressor trained for binary classification with the Calabi-Yau weights and the: (a) Random weights, (b) Transverse weights; predicting all the Calabi-Yau weights to be Calabi-Yau or not. The classification results are plotted with respect to the hypersurface's non-trivial Hodge numbers, showing perfect prediction above certain bounds for $h^{2,1}$ or $h^{1,1}$ respectively.}\label{fig:LR_misclass}
\end{figure}

However these accuracy scores are not at the level of usual successful machine learning investigations.
Hence to probe this further the misclassifications of the trained logistic regressor predicting on all the Calabi-Yau data were examined.
These results plotted against their respective $(h^{1,1},h^{2,1})$ values showed a clear correlation where pre-training against random data only misclassified at low $h^{2,1}$ values, and pre-training against transverse data only misclassified at low $h^{1,1}$.
This behaviour is shown in Figure \ref{fig:LR_misclass}.

Further investigations then binned the data according to the Hodge numbers, and trained/tested within each bin.
Interestingly, which weights took dominant importance flipped between the $h^{1,1}$ and $h^{2,1}$ binning investigations.

\section{Polytopes and Toric Geometry}\label{polytopes}
For local/non-compact CY $(d+1)$-folds, their information can be combinatorially encoded by convex polytopes on the lattice $N:=\mathbb{Z}^d$. Roughly speaking, one constructs the $(d+1)$-dimensional cone generated by the polytope. Then the maximal spectrum of the group algebra generated by its dual cone yields a Gorenstein singularity, which can be resolved to a local CY $(d+1)$-fold whose compact base is the Gorenstein Fano toric $d$-fold specified by the polytope. In other words, considering the fan of the polytope and the $d$-dimensional cones therein, we can construct a projective $d$-dimensional variety by gluing the affine varieties from these cones. In this section, we shall mainly focus on the polytopes corresponding to toric Fano varieties.
\begin{definition}
    A polytope $P$ is Fano if the origin $\bm{0}$ is strictly in its interior $\textup{int}(P)$ and all its vertices are primitive\footnote{A lattice point $\bm{v}=(v_1,\dots,v_d)$ is primitive if gcd$(v_i)=\pm1$.}. It is further called a canonical Fano polytope if $\textup{int}(P)\cap\mathbb{Z}^d=\{\bm{0}\}$.
\end{definition}
It is known that there are 16 toric Fano 2-folds with at worst canonical singularity which are precisely the reflexive polygons. For 3-folds, there are 674688 of them\footnote{In contrast, there are only 4319 reflexive polyhedra due to the extra criterion that not only is there a single interior point but also all bounding facets/edges are distance 1 to this point.} as classified in \cite{kasprzyk2010canonical}. We will then deal with those canonical Fano 3-folds in our machine learning problem while general Fano varieties are considered for $d=2$.

The set of vertices of a lattice polytope is a natural option as our input. However, as pointed out in \cite{Bao:2021ofk}, the vector of Pl\"ucker coordinates as a geometric invariant theory representation is a better choice for learning various properties of the polytope.
\begin{definition}
Let $P$ be a $d$-dimensional polytope with $n$ vertices.
This structure can be reformatted into an $d \times n$ matrix $V$ where each column corresponds to a vertex $\bm{v}_i$.
Consider the integer kernel of $V$ (aka grading) and take the maximal minors of $\ker(V)$; this gives a list of integers known as the Pl\"ucker coordinates, which we regard as a point in projective space.
\end{definition}
The Pl\"ucker coordinates reveal linear relations among vertices, and it is not hard to see that they are invariant under GL$(d,\mathbb{Z})$ transformations. However, this also gives rise to the ambiguity representing a unique polytope $P$. One may introduce the quotient gradings to further distinguish those polytopes. Here, we shall instead impose the condition that the vertices of $P$ generate the lattice $\mathbb{Z}^d$. This uniquely determines $P$.

Moreover, the initial ordering of the $n$ vertices leads to $n!$ equivalent ways of writing the Pl\"ucker coordinates. This is due to the symmetric group $\mathfrak{S}_n$ permuting the vertices. Such redundancy is irrelevant for Pl\"ucker coordinates representing a polytope, but this can be used to enhance our datasets. We will choose 3 and 10 distinct Pl\"ucker coordinate orderings for each polygon and polyhedron respectively in our data.

\subsection{Volumes and Dual Volumes}\label{voldualvol}
\begin{figure}[h]
	\centering
    \begin{subfigure}{3.5cm}
    	\centering
    	\includegraphics[width=3.5cm]{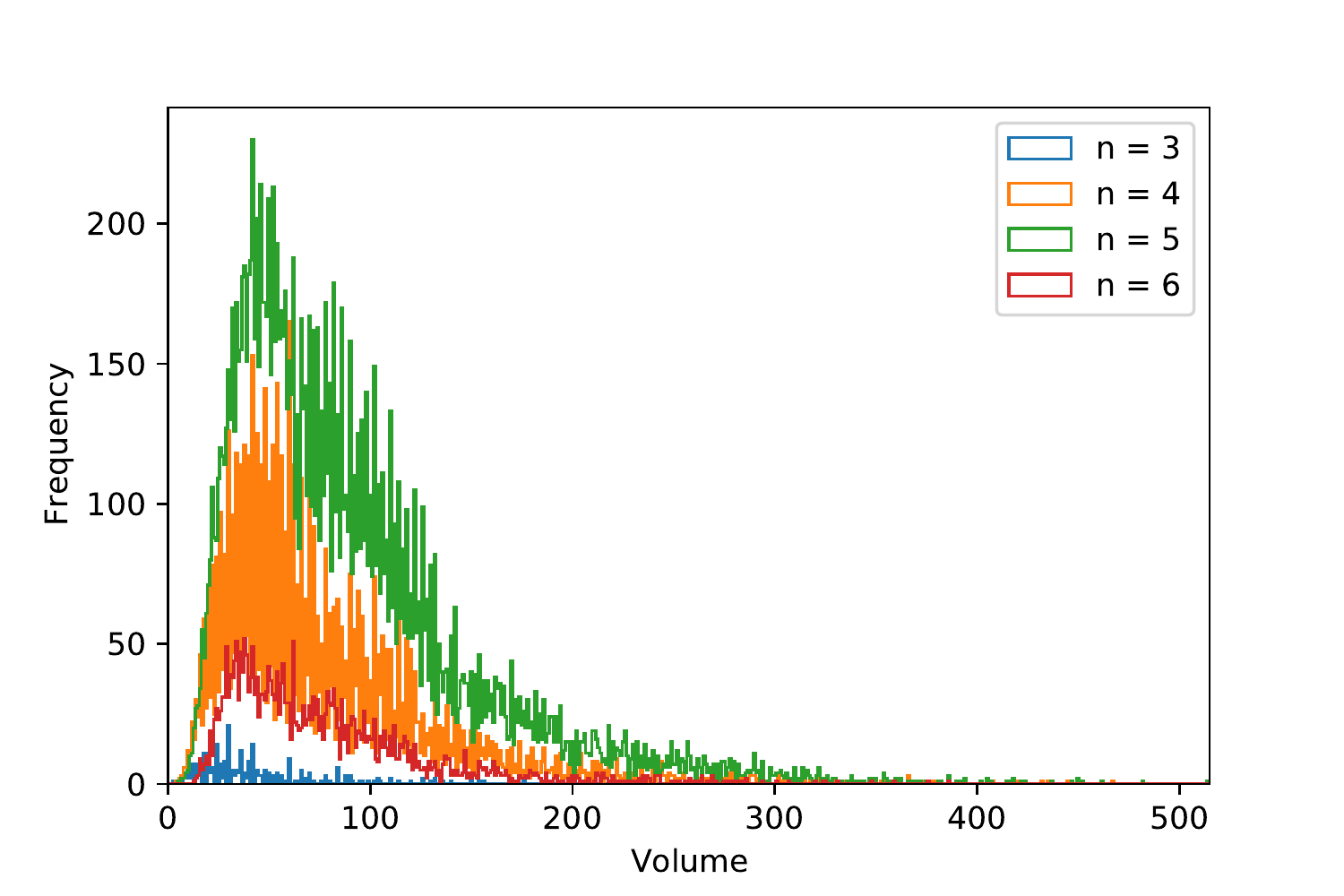}
    	\caption{}\label{polygon_Vol}
    \end{subfigure}
    \begin{subfigure}{3.5cm}
    	\centering
    	\includegraphics[width=3.5cm]{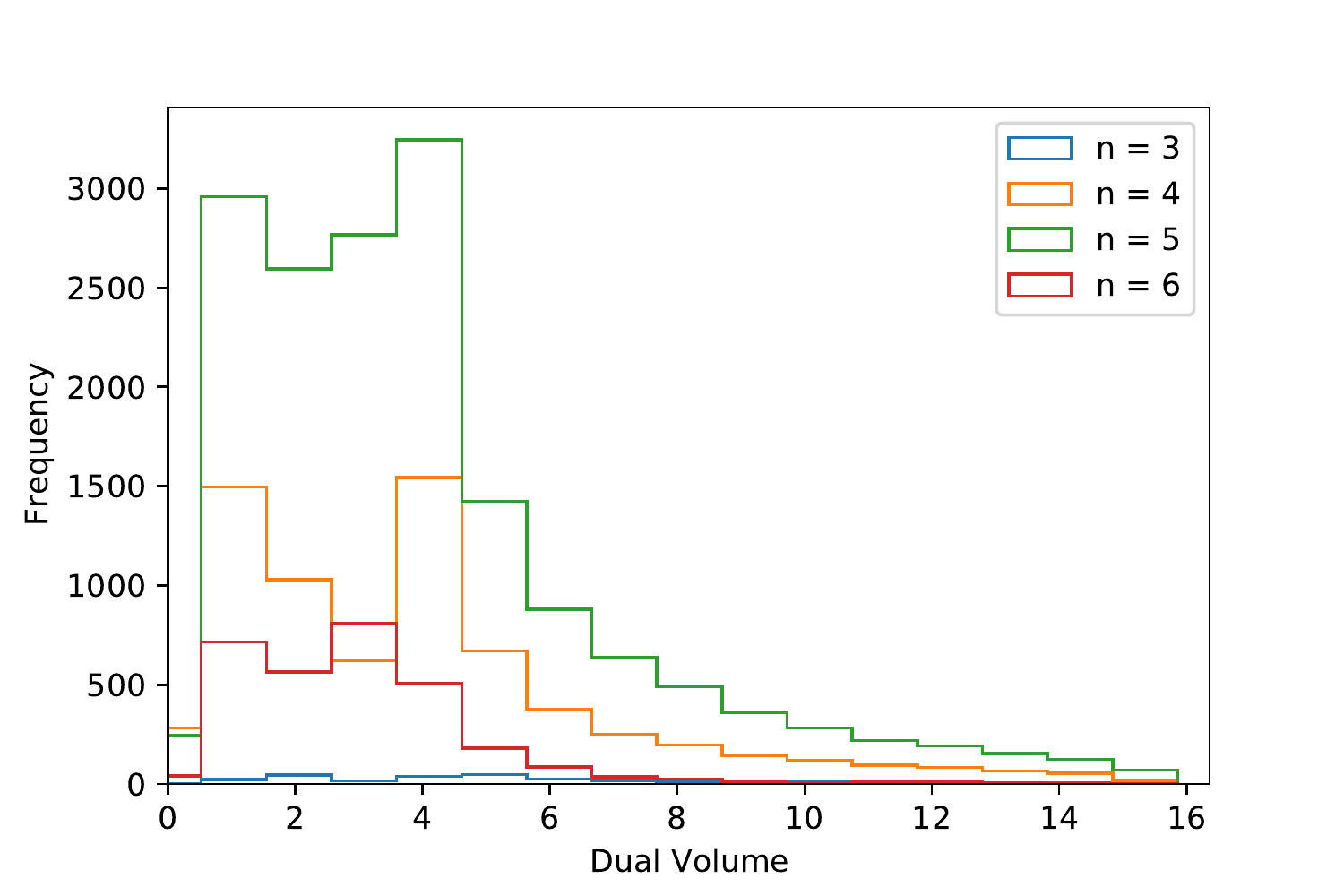}
    	\caption{}\label{polygon_DualVol}
    \end{subfigure}
    \begin{subfigure}{3.5cm}
    	\centering
    	\includegraphics[width=3cm]{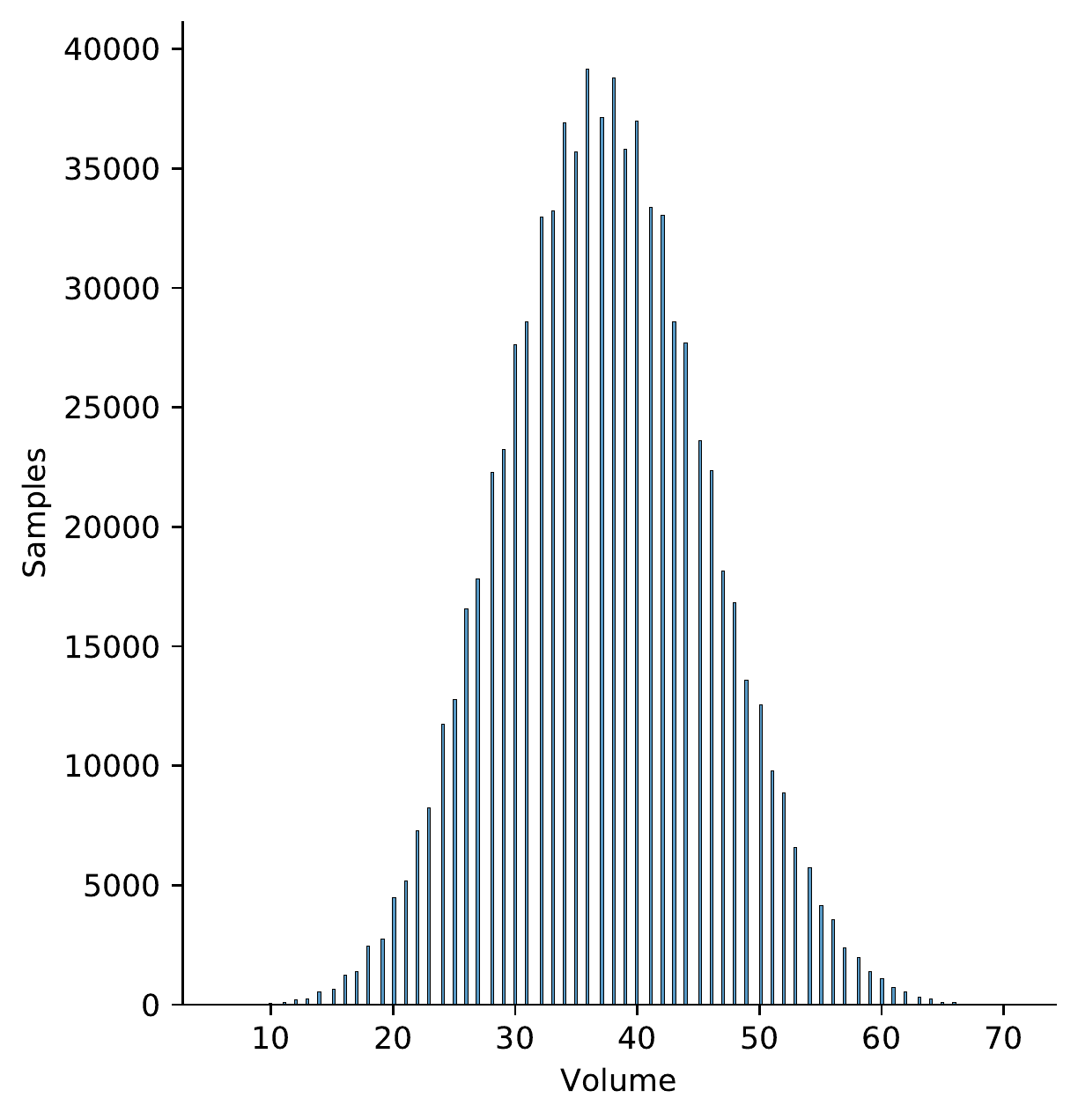}
    	\caption{}\label{polygon_Vol}
    \end{subfigure}
    \begin{subfigure}{3.5cm}
    	\centering
    	\includegraphics[width=3cm]{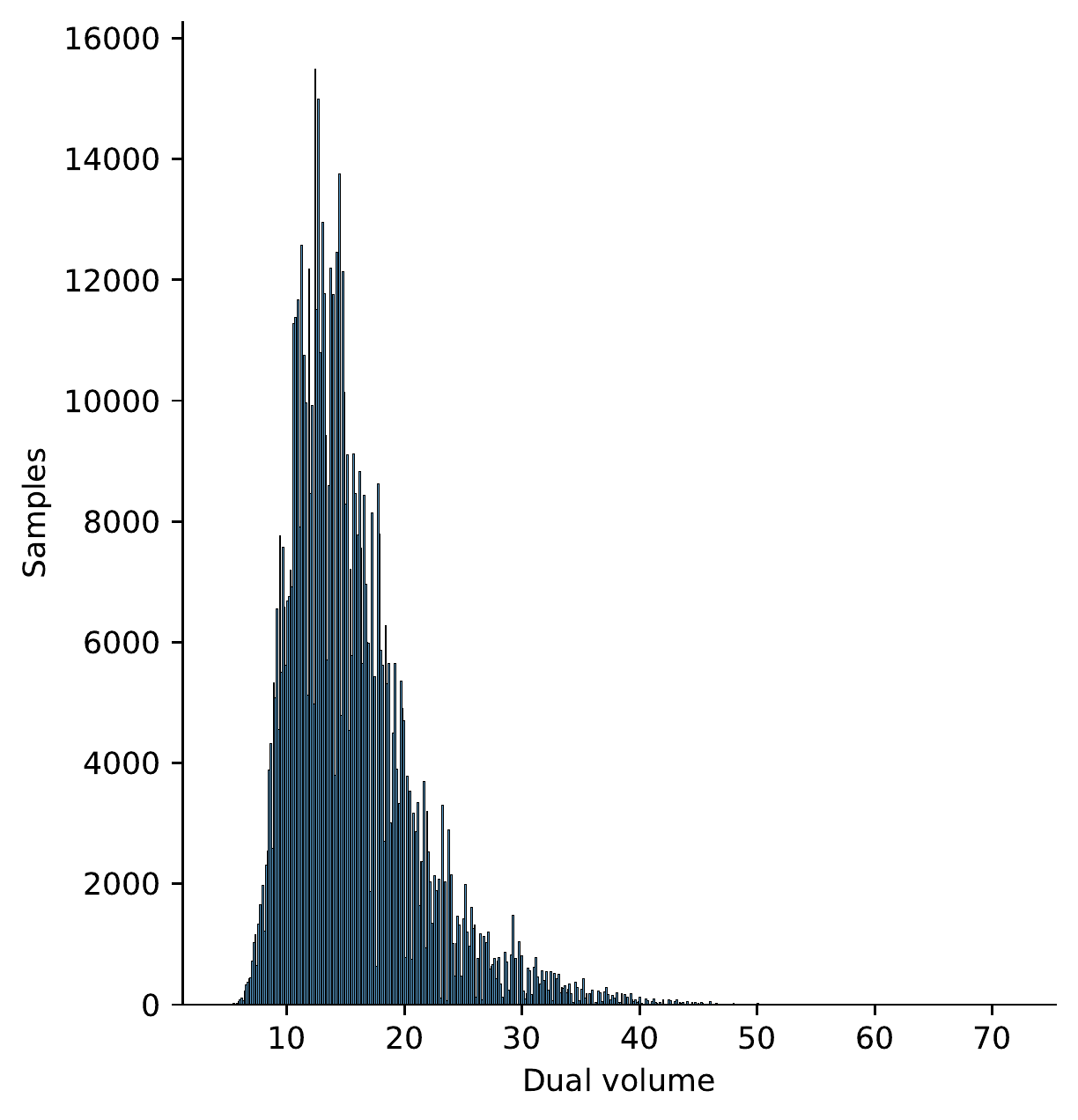}
    	\caption{}\label{polygon_DualVol}
    \end{subfigure}
    \caption{Distributions of the (a) volumes and (b) dual volumes of 2d polygons and distributions of the (c) volumes and (d) dual volumes of 3d polyhedra.}\label{poly_dists}
\end{figure}

Give a polytope, two straightforward quantities one can obtain are its volume and dual volume (i.e., the volume of its dual polytope). In particular, we take the normalized volume where each simplex has volume 1. For $d=2$, we generate random Fano polygons with the numbers of vertices $n\in\{3,4,5,6\}$. The volumes and dual volumes take values within the range $[3,514]$ and $[0.21,15.37]$ respectively. Their distributions are shown in Figure \ref{poly_dists} (a, b). For each $n$, a network is trained separately.

We can perform the similar experiment for 3d polytopes as well. Since there is sufficient data, the randomly chosen polyhedra are restricted to be canonical Fano and we will not train different networks for different $n$ as in the 2d case. The distributions of data are given in Figure \ref{poly_dists} (c,d).

Here, we apply MLP to this regression problem\footnote{The detailed structures of the networks can be found in \cite{Bao:2021ofk}. Slight changes in the networks should not affect our results in mathematical problems \cite{He:2021oav}.}. As the Pl\"ucker coordinates for polytopes can have different lengths, we find that padding zeros to the input vectors would give best performance. The results are summarized in Table \ref{tablevoldualvol}. We may also use the Pl\"ucker coordinates as input to predict other properties of the polytopes. More details can be found in \cite{Bao:2021ofk}.

\begin{table}[h]
\centering
\begin{tabular}{V{4}cV{4}c|c|c|cV{4}}
\hlineB{4}
\begin{tabular}[c]{@{}c@{}}Polygons\\ (frequency)\end{tabular}   & \begin{tabular}[c]{@{}c@{}}Triangles\\ (277)\end{tabular} & \begin{tabular}[c]{@{}c@{}}Quadrilaterals\\ (7041)\end{tabular} & \begin{tabular}[c]{@{}c@{}}Pentagons\\ (16637)\end{tabular} & \begin{tabular}[c]{@{}c@{}}Hexagons\\ (3003)\end{tabular} \\ \hlineB{3}
Output & \multicolumn{4}{cV{4}}{Volume} \\ \hline
MAE & 0.209 & 0.625 & 1.051 & 3.359 \\ \hline
Accuracy & 1.000 & 1.000 & 1.000 & 0.997  \\ \hlineB{3}
Output  & \multicolumn{4}{cV{4}}{Dual Volume} \\ \hline
MAE & 1.181 & 0.642 & 0.818 & 0.941 \\ \hline
Accuracy & 0.501 & 0.754 & 0.638 & 0.557 \\ \hlineB{4}
\begin{tabular}[c]{@{}c@{}}Polyhedra\\ (frequency)\end{tabular}  & \multicolumn{4}{cV{4}}{\begin{tabular}[c]{@{}c@{}}Toric Canonical Fano Polyhedra\\ (780000)\end{tabular}} \\ \hlineB{3}
Output & \multicolumn{2}{cV{1}}{Volume} & \multicolumn{2}{cV{4}}{Dual Volume} \\ \hline
MAE & \multicolumn{2}{cV{1}}{1.680} & \multicolumn{2}{cV{4}}{2.590} \\ \hline
Accuracy & \multicolumn{2}{cV{1}}{0.936} & \multicolumn{2}{cV{4}}{0.890} \\ \hlineB{4}
\end{tabular}
\caption{The machine learning results for volumes and dual volumes. The ``Accuracy'' for polygons is the accuracy$\pm0.05\times\text{range}$. Likewise, for those involving polyhedra, the ``Accuracy'' is the accuracy$\pm4$. The frequency stands for the number of samples trained in each model.}\label{tablevoldualvol}
\end{table}



\subsection{Data Projection and Visualization}\label{projpolytope}
\begin{figure}[h]
	\centering
	\begin{subfigure}{3.5cm}
	\includegraphics[width=3.5cm]{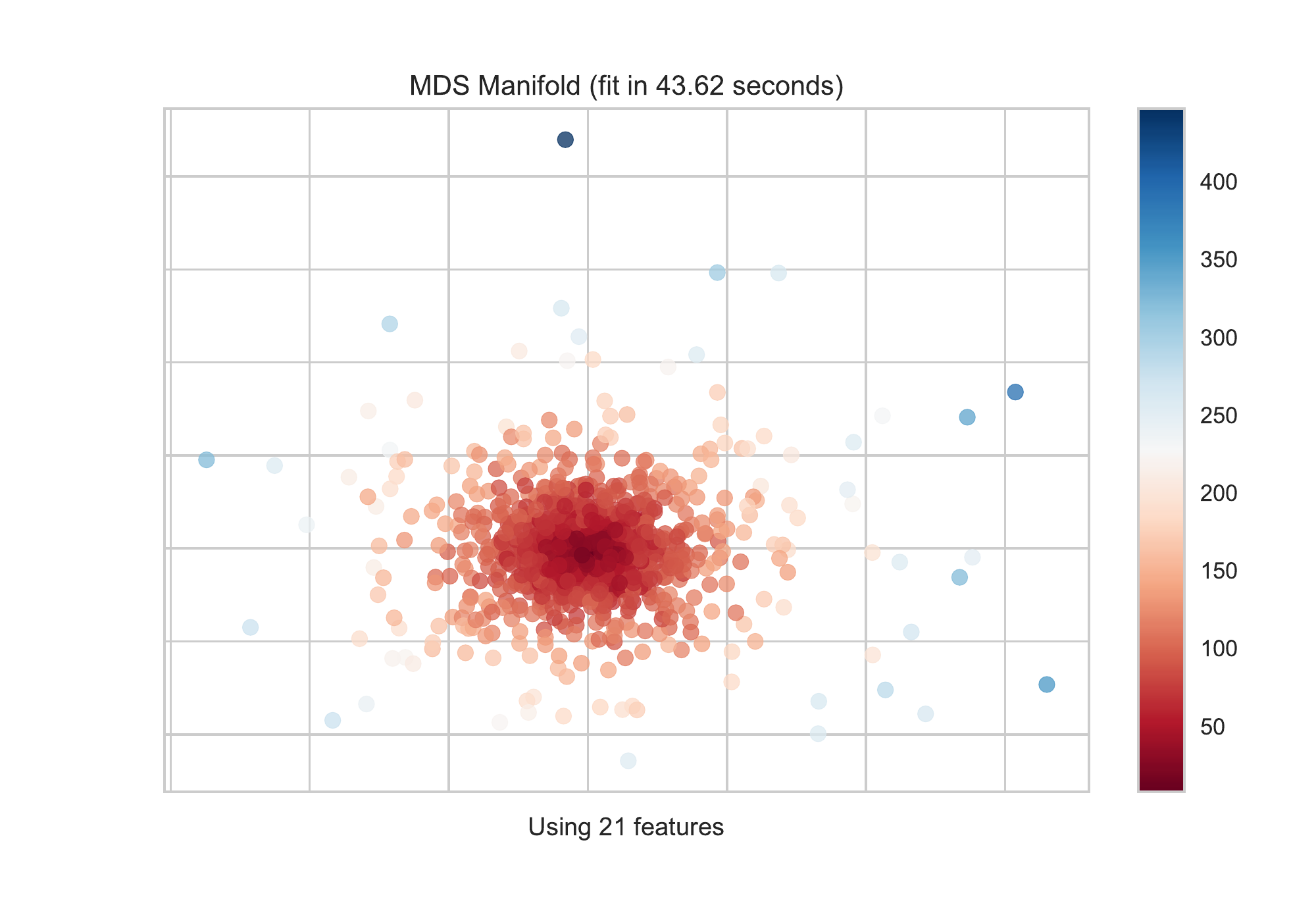}
    \caption{}
	\end{subfigure}
	\begin{subfigure}{3.5cm}
	\includegraphics[width=3.5cm]{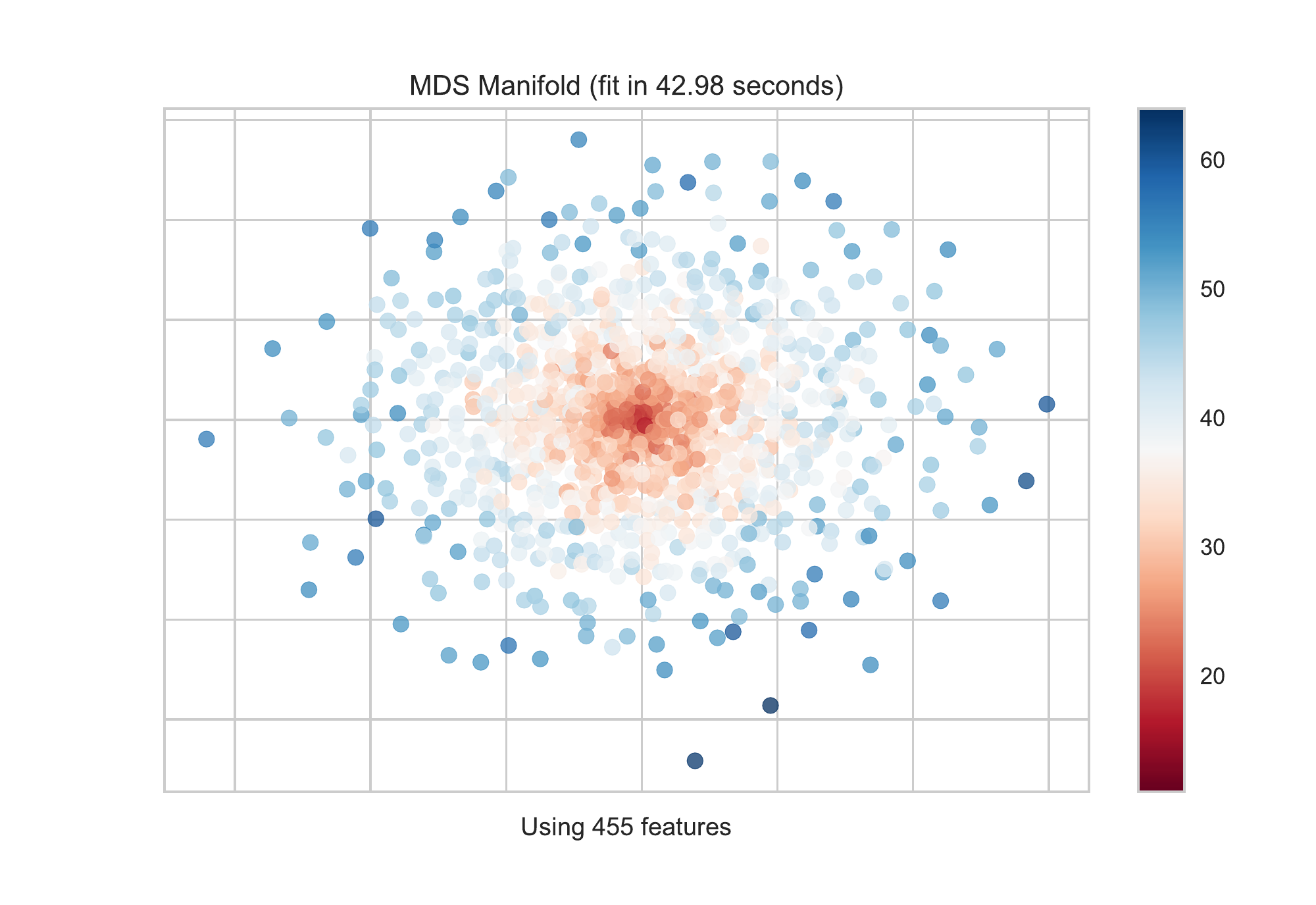}
    \caption{}
	\end{subfigure}
	\begin{subfigure}{3.5cm}
	\includegraphics[width=3.5cm]{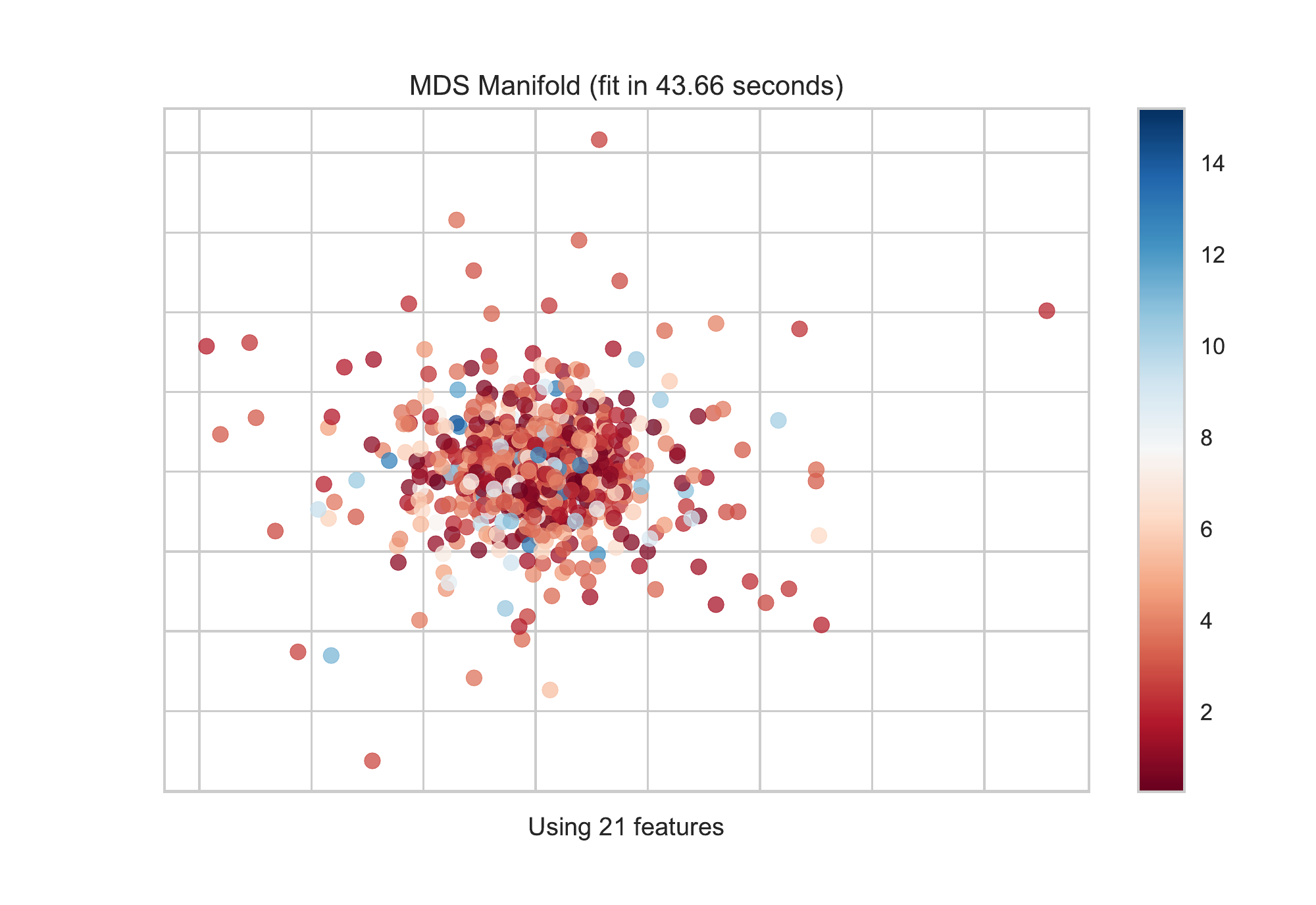}
    \caption{}
	\end{subfigure}
	\begin{subfigure}{3.5cm}
	\includegraphics[width=3.5cm]{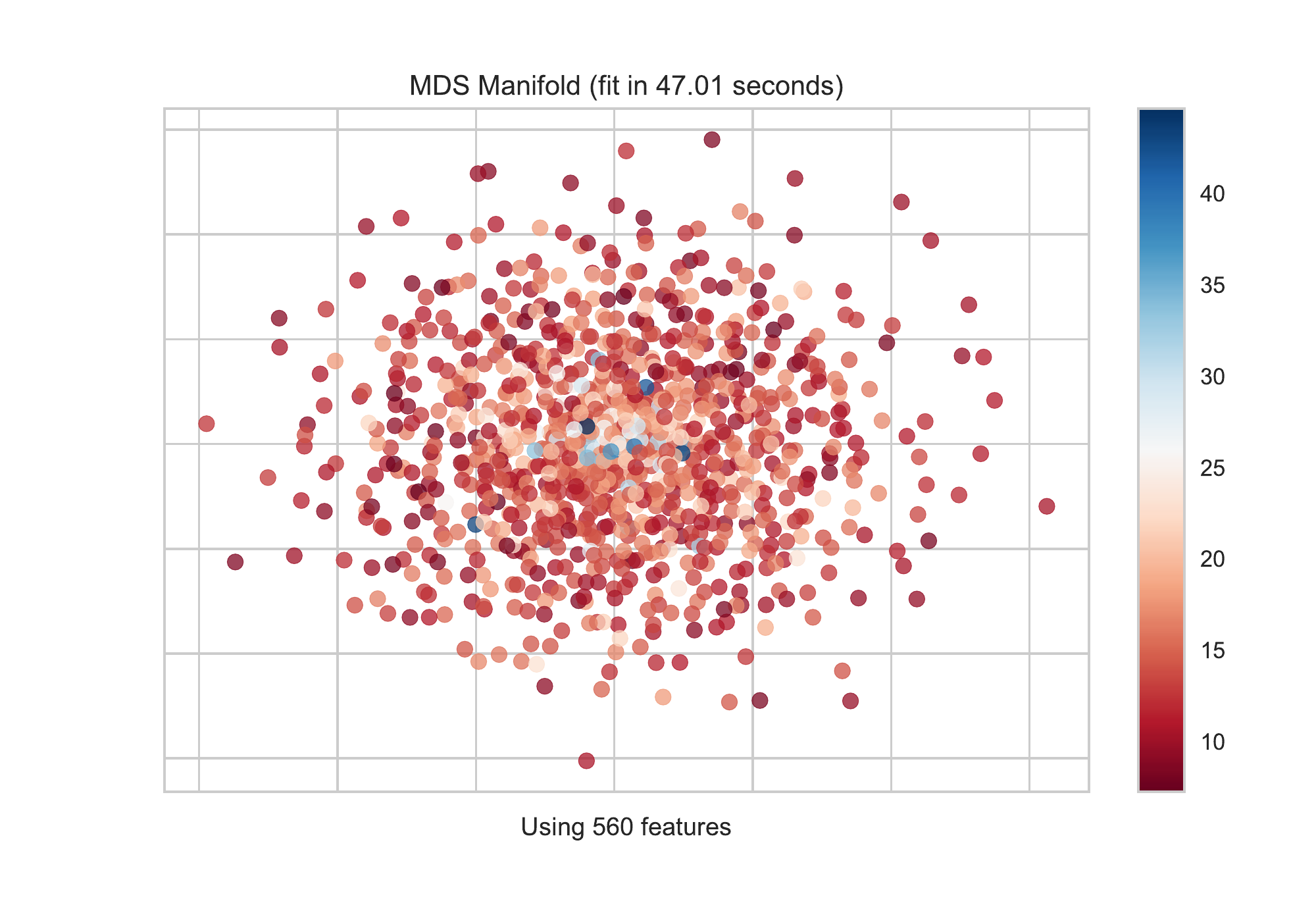}
    \caption{}
	\end{subfigure}
	\caption{The MDS projections of Pl\"ucker coordinates. The volumes are given in (a) polygons and (b) for polyhedra. The dual volumes are given in (c) polygons and (d) for polyhedra. Different colours of data points indicate different (dual) volumes. Here, we use 1000 random samples for each plot as an illustration.}\label{voldualvolmds}
\end{figure}

To study the potential structures captured by the machine, one may try to project the input data to lower dimensions. A handy way to perform this unsupervised learning technique is the so-called manifold learning \cite{bengfort_yellowbrick_2019}. For instance, the multi-dimensional scaling (MDS)\footnote{Readers are referred to \cite{borg2005modern} for an introduction to MDS.} projects a Pl\"ucker input vector $\bm{p}=(p^0:\dots:p^{l-1})$ of length $l$ to some vector $x=(x^0,\dots,x^{k-1})$ with $k<l$ by minimizing the cost function (aka stress)
\begin{equation}
    S=\left(\sum_{i<j\leq N}(D_{ij}-d_{ij})^2\right)^{1/2},
\end{equation}
where $N$ is the number of samples and $D_{ij}$ ($d_{ij}$) denotes the Euclidean distance between $\bm{p}_i$ ($\bm{x}_i$) and $\bm{p}_j$ ($\bm{x}_j$). The MDS projections for polygons and polyhedra are shown in Figure \ref{voldualvolmds} (a, b) with the hue based on volumes.

As we can see, the vectors $\bm{x}$ are scattered around a centre point. The closer $\bm{x}$ is to the centre, the smaller volume the corresponding polytope would have. Since this is a mathematical fact, the number of samples should not affect the results and minimizing $S$ is therefore equivalent to the goal of finding $\bm{x}$ such that $D_{ij}-d_{ij}\approx0$ for any $i,j$. As the positions of the data points with different volumes mainly depend on the distances from the centre point in Figure \ref{voldualvolmds} (a, b), 1-dimensional projections are actually sufficient as shown in \cite{Bao:2021ofk}. We are using the 2-dimensional plots here for later comparison with dual volumes. From the MDS projections, it seems to indicate a map $f:\bm{p}\mapsto x$ such that $\text{Vol}\sim x+1$. This should be in line with the fact that $\text{Vol}=\sum\limits_{\alpha\in\mathcal{V}}p^{\alpha}$ where $\mathcal{V}\subseteq\{0,\dots,l-1\}$. More detailed analysis can be found in \cite{Bao:2021ofk}.

Now let us apply MDS to dual volumes. As visualized in Figure \ref{voldualvolmds} (c, d), the points with different colours are distributed more randomly. This also explains why the machine learning performance on dual volumes is inferior to the one on volumes.

\section{Hilbert Series}\label{HS}
Hilbert series (HS) are important tools in modern geometry. Their use in physics centres in supersymmetric quantum field theories where they are useful for BPS operator counting \cite{Feng:2007ur}, as well as in more general string theory where they provide geometric information about a theory's moduli space \cite{Mehta:2012wk}.

Mathematically, HS encode embedding-specific information about a complex projective variety $X$, embedded in $\mathbb{P}_\mathbb{C}(p_0^{q_0}, \ldots, p_s^{q_s})$, denoted where weight $p_i$ occurs $q_i$ times. 
The embedding induces a grading on its coordinate ring $R=\mathbb{C}[X_0,\ldots,X_k]/I$, such that $R=\bigoplus_{i\ge0} R_i$; for variables $X_i$ and $I$ the homogeneous ideal generated by the polynomials defining the variety \cite{Dolga}.
Now the HS is the generating function for the dimensions of the graded pieces of $R$,
\begin{equation}
    H(t; X) = \sum\limits_{i=0}^{\infty} (\dim_{\CC} R_i) t^i \;;
\end{equation}
such that dim$_{\mathbb{C}}R_i$ is also the number of independent degree $i$ polynomials on the variety.

Via the Hilbert-Serre theorem \cite{AtiyahMacdonald}, the HS can be written in two forms:
\begin{equation}
    (1) \quad H(t; X)=\frac{P(t)}{\prod\limits_{i=0}^s(1-t^{p_i})^{q_i}},\quad(2) \quad H(t; X)= \frac{\tilde{P}(t)}{(1-t^j)^{dim+1}},\label{hs_form}
\end{equation}
for polynomials $P$ and $\tilde{P}$ such that the variety has Gorenstein index $j$ and dimension $dim$.
In addition if the numerator polynomial $\tilde{P}$ is palindromic the variety is Gorenstein in nature \cite{Stanley1978}.

The work in \cite{Bao:2021auj,Hirst:2022qqr} initiates the application of machine learning techniques to these objects, with the aim that from the first terms of the series' Taylor expansions one can extract information about the underlying variety from the HS closed forms from \eqref{hs_form}.
With this information one can then compute all higher terms, and hence information about any order BPS operators from just information on the first few, as well information about the physical theory's moduli space.

\subsection{Gathering HS Data}
The data used in the investigations included HS associated to algebraic varieties, retrieved from the~GRDB~\cite{grdb,fanodata}. These used a database of candidate~HS conjecturally associated to three-dimensional~$\QQ$-Fano varieties with Fano index one, as constructed in~\cite{ABR:Fano,BK22}.
In addition data was also generated by sampling the hyperparameters of the HS form \eqref{hs_form} from distributions fitted from the GRDB data; ensuring that some physical conditions were met to make the data representative and that there was no overlap with the GRDB data.

For both HS datatypes, which we call `real' and `fake' respectively, the HS closed forms are Taylor expanded to provide coefficient lists. 
The coefficient lists are hence vectors of integers, and make up the ML input for the investigations carried out.
Two coefficient vectors were used as inputs, one from the start of the series (coefficients 0-100) and one from deeper into it (1000-1009).
Using the higher order terms in a separate vector as well allows one to assess the importance of orbifold points which take effect deeper in the series leading to the coefficient growth rate being more stable.

\subsection{Learning Geometric Properties}
In each of the five investigations performed with this data different properties of the HS were learnt.
NN regressors were used to predict the vector of embedding weights $(p_i)$, whilst classifiers learnt the Gorenstein index $j$ and dimension $dim$ directly.
Further to this three binary classification investigations were performed to evaluate whether HS were Gorenstein, were complete intersections, or whether they were from the GRDB/generated by us.

\paragraph{Architectures} 
For all investigations the NNs used had 4 dense layers of 1024 neurons and 0.05 dropout, each with ReLU activation. 
Training was in batches of 32 for 20 epochs using the Adam optimiser to minimise either the log(cosh) or cross-entropy loss functions for regression or classification respectively.
5-fold cross-validation was used to provide confidence for the metrics used to evaluate the learning. 
For regression this was assessed with the MSE metric, and for classification accuracy and MCC were used.

\paragraph{Results}
Results for each of the investigations are given in Tables \ref{HS_results} and \ref{HS_resultsbinary}. 
The regressor results show that the weights of the ambient space could be learnt to the nearest integer, performing better with higher orders.
The geometric parameters were learnt with exceptional accuracy also, but conversely performed better using lower order coefficients.

The binary classification results in Table \ref{HS_resultsbinary} show the Gorenstein and complete intersection properties could be confidently detected from higher orders, whilst the true GRDB vs fake generated data was trivially distinguished, as also corroborated by PCA, this indicates how difficult it is to generate representative HS data.

\begin{table}[t]
\centering
\begin{tabular}{|c|c|cc|}
\hline
\multirow{2}{*}{Investigation}      & \multirow{2}{*}{\begin{tabular}[c]{@{}c@{}}Orders\\ of Input\end{tabular}} & \multicolumn{2}{c|}{Performance Measure}   \\ \cline{3-4} 
    &    & \multicolumn{2}{c|}{MAE}   \\ \hline
\multirow{2}{*}{Weights $(p_i)$}    & 0-100      & \multicolumn{2}{c|}{$1.94 \pm 0.11$}       \\ \cline{2-4} 
    & 1000-1009  & \multicolumn{2}{c|}{$1.04 \pm 0.12$}       \\ \hline
-   & -  & \multicolumn{1}{c|}{Accuracy}  & MCC       \\ \hline
\multirow{2}{*}{\begin{tabular}[c]{@{}c@{}}Gorenstein\\ Index $j$\end{tabular}} & 0-100      & \multicolumn{1}{c|}{$0.934 \pm 0.008$} & $0.916 \pm 0.010$ \\ \cline{2-4} 
    & 1000-1009  & \multicolumn{1}{c|}{$0.780 \pm 0.018$} & $0.727 \pm 0.022$ \\ \hline
\multirow{2}{*}{\begin{tabular}[c]{@{}c@{}}Dimension\\ $dim$\end{tabular}}  & 0-100      & \multicolumn{1}{c|}{$0.995 \pm 0.005$} & $0.993 \pm 0.006$ \\ \cline{2-4} 
    & 1000-1009  & \multicolumn{1}{c|}{$0.865 \pm 0.024$} & $0.822 \pm 0.031$ \\ \hline
\end{tabular}
\caption{Results for learning parameters of the full Hilbert series from vectors of coefficients at specified orders. Learning embedding weights of Form (1) in \eqref{hs_form} used NN Regressors, whilst learning Gorenstein index and dimension from Form (2) in \eqref{hs_form} used NN classifiers.}\label{HS_results}
\end{table}

\begin{table}[tb]
\centering
\begin{tabular}{|c|c|cc|}
\hline
\multirow{2}{*}{Investigation}      & \multirow{2}{*}{\begin{tabular}[c]{@{}c@{}}Orders\\ of Input\end{tabular}} & \multicolumn{2}{c|}{Performance Measures}  \\ \cline{3-4} 
&   & \multicolumn{1}{c|}{Accuracy}  & MCC       \\ \hline
\multirow{2}{*}{Gorenstein} & 0-100 & \multicolumn{1}{c|}{$0.844 \pm 0.087$} & $0.717 \pm 0.155$ \\ \cline{2-4} 
& 1000-1009     & \multicolumn{1}{c|}{$0.954 \pm 0.043$} & $0.919 \pm 0.073$ \\ \hline
\multirow{2}{*}{\begin{tabular}[c]{@{}c@{}}Complete\\ Intersection\end{tabular}} & 0-100 & \multicolumn{1}{c|}{$0.762 \pm 0.010$} & $0.544 \pm 0.030$ \\ \cline{2-4} 
& 0-300 & \multicolumn{1}{c|}{$0.951 \pm 0.005$} & $0.902 \pm 0.010$ \\ \hline
\multirow{2}{*}{GRDB}       & 0-100 & \multicolumn{1}{c|}{$1.000 \pm 0.000$} & $1.000 \pm 0.000$ \\ \cline{2-4} 
& 1000-1009     & \multicolumn{1}{c|}{$1.000 \pm 0.000$} & $1.000 \pm 0.000$ \\ \hline
\end{tabular}
\caption{Binary classification results for distinguishing whether input vectors of HS coefficients of specified orders come from full HS  which represent Gorenstein varieties, complete intersections, or whether the HS was from the GRDB or generated as `fake' data.}\label{HS_resultsbinary}
\end{table}

\section{The Genus of Amoebae}\label{amoeba}
Given a Newton polynomial $P(z,w)$, its amoeba is the set\footnote{Amoebae can certainly be defined for any Laurent polynomial $P(z_1,\dots,z_r)$. Nevertheless, we shall focus on $r=2$ in this section.}
\begin{equation}
    \mathcal{A}_P:=\{(\log|z|,\log|w|)\in\mathbb{R}^2:P(z,w)=0\}.
\end{equation}
Amoeba is an important concept in tropical geometry \cite{mikhalkin2004amoebas,gelfand2008discriminants,itenberg2009tropical} and has many applications in physics \cite{Kenyon:2003uj,Feng:2005gw,Ooguri:2009ijd,Zahabi:2020hwu,Bao:2021fjd}. Roughly speaking, an amoeba is the thickening of the dual web of its associated toric diagram. It is often not easy to determine the boundary of an amoeba for a generic Newton polynomial. In \cite{purbhoo2008nullstellensatz}, the idea of lopsidedness was applied to find the boundaries of any amoebae. We say a list of positive numbers is lopsided if one of the numbers is greater than the sum of all the others. In particular, a list $P\{z,w\}$ can be constructed from $P(z,w)$ as $P\{z,w\}=\{|m_i(z,w)|\}$ where $m_i(z,w)$ are the monomials of $P(z,w)$. This yields the so-called lopsided amoeba $\mathcal{LA}_P=\{(z,w)|P\{z,w\}\text{ is not lopsided}\}$. This provides an approximation of the amoeba and its boundary in the sense that $\mathcal{LA}_P\supseteq\mathcal{A}_P$. To get better approximations, we may consider the cyclic resultant of $P(z,w)$, or equivalently,
\begin{equation}
    \widetilde{P}_n(z,w):=\prod_{k_1=0}^{n-1}\prod_{k_2=0}^{n-1}P\left(\text{e}^{2\pi ik_1}z,\text{e}^{2\pi ik_2}w\right).
\end{equation}
Then $\mathcal{LA}_{\widetilde{P}_n}$ converges uniformly to $\mathcal{A}_P$ as $n\rightarrow\infty$. In other words, we can get better approximations of the amoeba by taking higher level $n$.

As a result, the complementary regions of an amoeba on $\mathbb{R}^2$ would also be determined once we know its boundary. In particular, the number of bounded complementary regions is called the genus of the amoeba. As a matter of fact, the maximal possible genus is equal to the number of internal points of the corresponding lattice polygon. Machine learning techniques were used in \cite{Bao:2021olg} to find the genus of an amoeba when varying the coefficients of $P(z,w)$.

Let us illustrate this with one of the simplest examples, that is, $F_0$ whose Newton polynomial reads $P=c_1z+c_2w+c_3z^{-1}+c_4w^{-1}+c_5$. It is not hard to see that the associated polygon has one single internal point. Therefore, the genus $g$ is either 0 or 1, and we simply have a binary classification problem. Moreover, we shall mainly focus on the lopsided amoeba $\mathcal{LA}_P$, that is, the approximation at $n=1$.

\subsection{Data Projection and Visualization}\label{projamoeba}
Using the coefficients $\{c_1,\dots,c_5\}$ (sampled $c_{1,2,3,4}\in[-5,5]$, $c_5\in[-20,20]$) as input and $g$ as output, a simple MLP can easily reach over 0.95 accuracy with fewer than 2000 training samples. We will discuss the structure of the network in the next subsection while here let us first apply some unsupervised techniques to understand what information we might be able to extract from the machine.

\begin{figure}[h]
    \centering
    \begin{subfigure}{4.5cm}
         \centering
         \includegraphics[width=4.5cm]{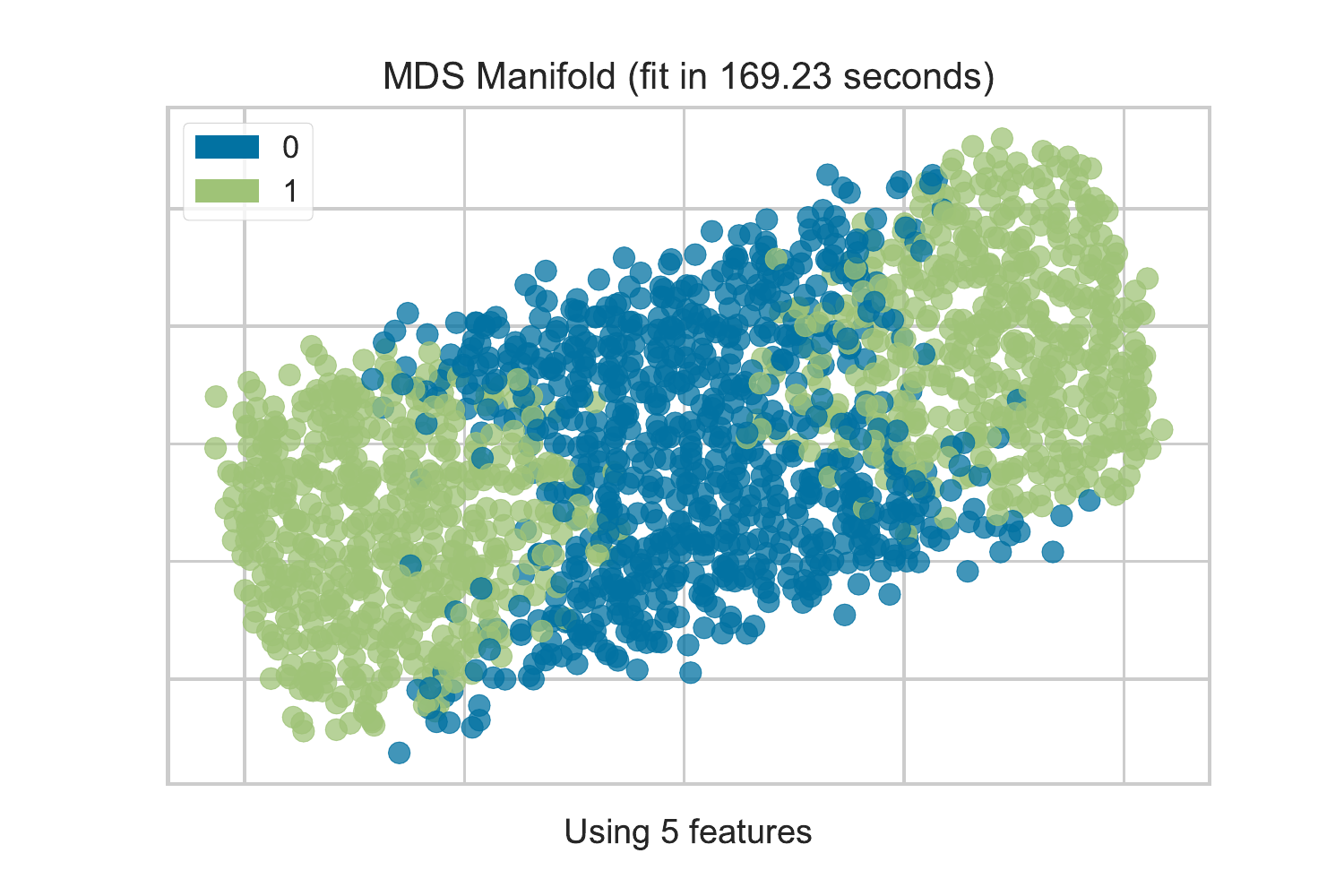}
         \caption{}
     \end{subfigure}
     \begin{subfigure}{4.5cm}
         \centering
         \includegraphics[width=4.5cm]{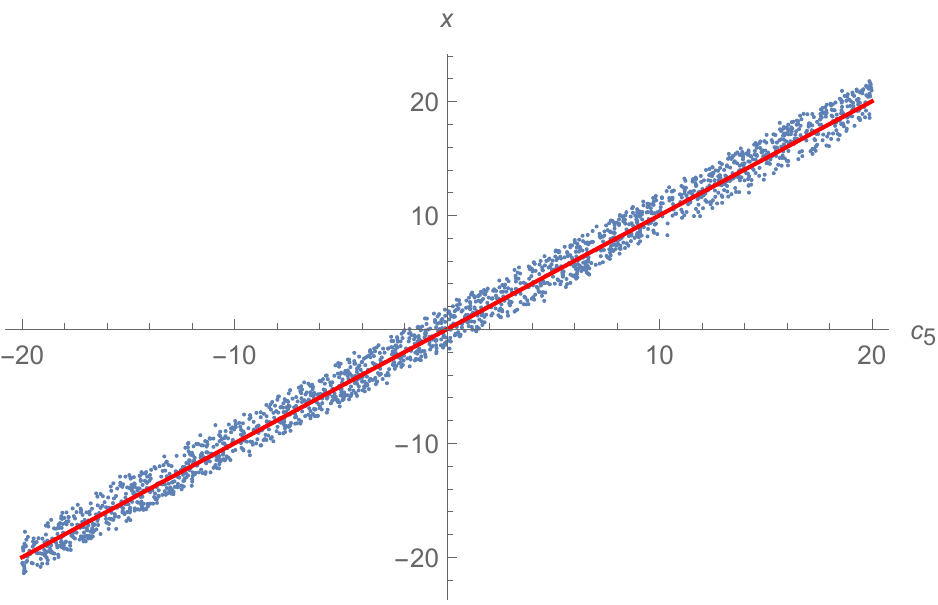}
         \caption{}
     \end{subfigure}
     \begin{subfigure}{4.5cm}
         \centering
         \includegraphics[width=4.5cm]{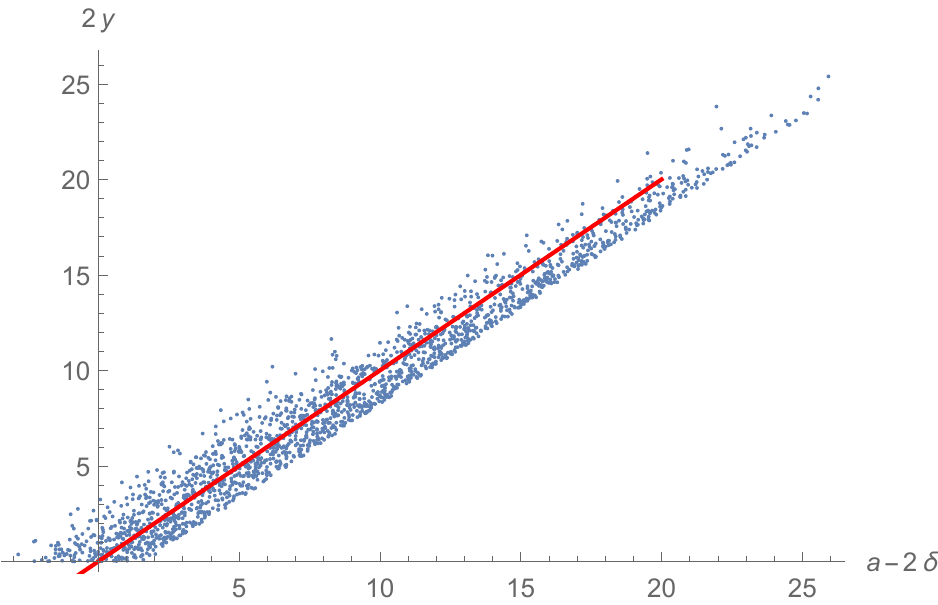}
         \caption{}
     \end{subfigure}
    \caption{The MDS manifold projection $(c_1,\dots,c_5)\mapsto(x,y)$ and the analysis of $(x,y)$.}\label{F0mds}
\end{figure}

To analyze relevant features of the problem, let us again apply MDS projection as visualized in Figure \ref{F0mds}(a).
As we can see, the blue ($g=0$) and green ($g=1$) points have a rather clear separation. To understand this scattering plot, let us analyze the two coordinates $x$ and $y$ separately.

In Figure \ref{F0mds}(b), we plot $x$ against $c_5$. The red line $x=c_5$ shows a good fit for the points. We may therefore conclude that one of the coordinates in the projection is simply the coefficient $c_5$. It is then natural to expect that $y$ would be transformed from the remaining four coefficients. By curve fitting, we find that $|y_\text{fit}|=0.141|c_1c_3|+0.166|c_2c_4|+2.411$. Astute readers may have already found that this is very close to the approximation of square roots. Indeed, let $a:=2|c_1c_3|^{1/2}+2|c_2c_4|^{1/2}$. Then it can be approximated as $a/2\approx0.1|c_1c_3|+1.2+0.1|c_2c_4|+1.2$. In Figure \ref{F0mds}(c), we plot $2|y|$ versus $a-2\delta$ where $\delta=|y_\text{fit}|-|y|$. Again, we find that the points scatter along the 45-degree line crossing the origin. Hence, the other transformed coordinate in the data projection actually indicates $|y|\approx a/2$.

Therefore, the genus should be encoded by two parameters: $c_5$ and $a$. Indeed, applying the lopsidedness condition, we find that $g=0$ if and only if $|c_5|\leq a$. This in fact agrees with the scattering plot in Figure \ref{F0mds}(a) where the separation of blue and green regions, albeit not perfectly precise, approximates the lines $|y|=|x|$.

\begin{figure}[h]
	\centering
	\begin{subfigure}{3.5cm}
		\centering
		\includegraphics[width=3.5cm]{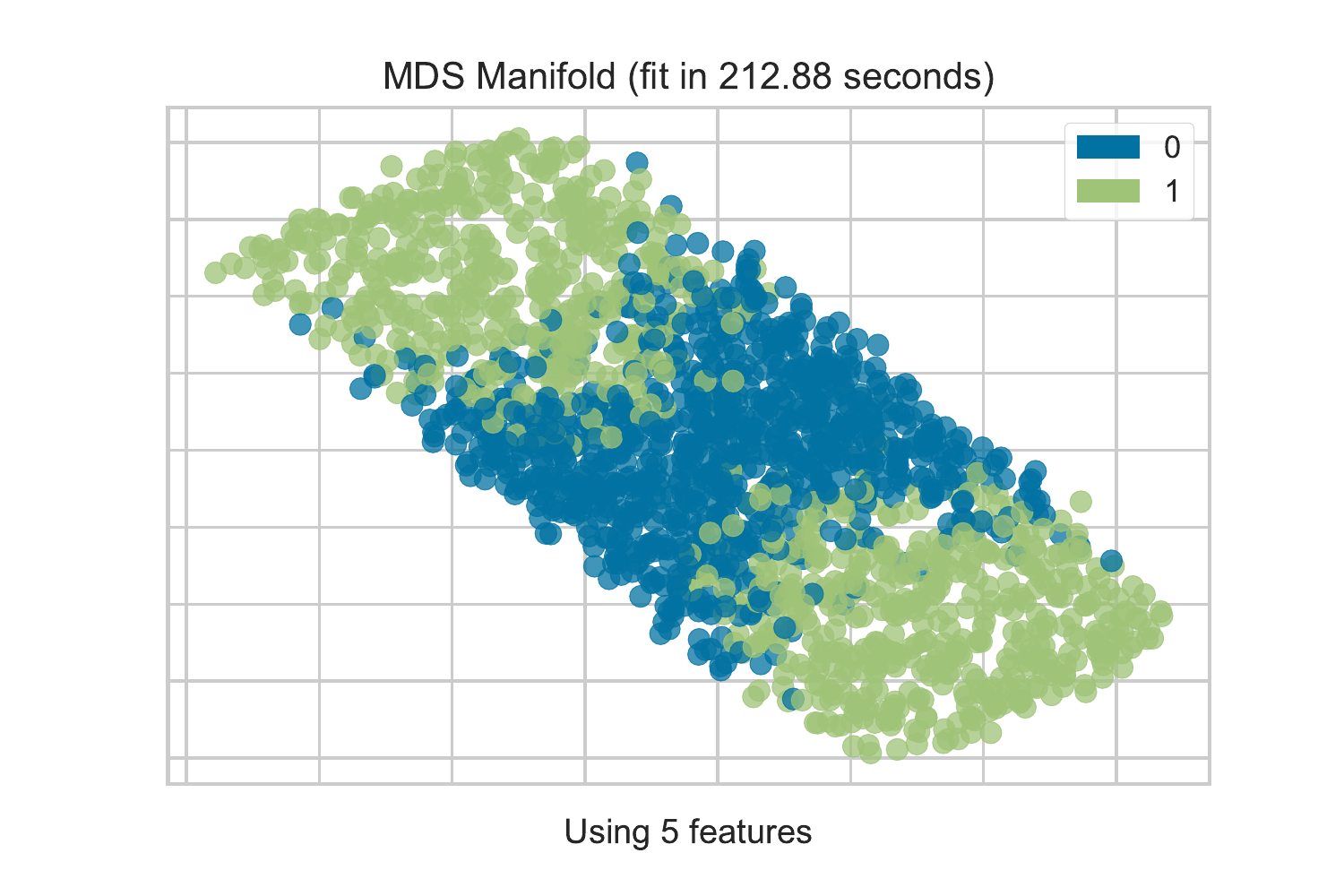}
		\caption{}
	\end{subfigure}
    \begin{subfigure}{3.5cm}
    	\centering
    	\includegraphics[width=3.5cm]{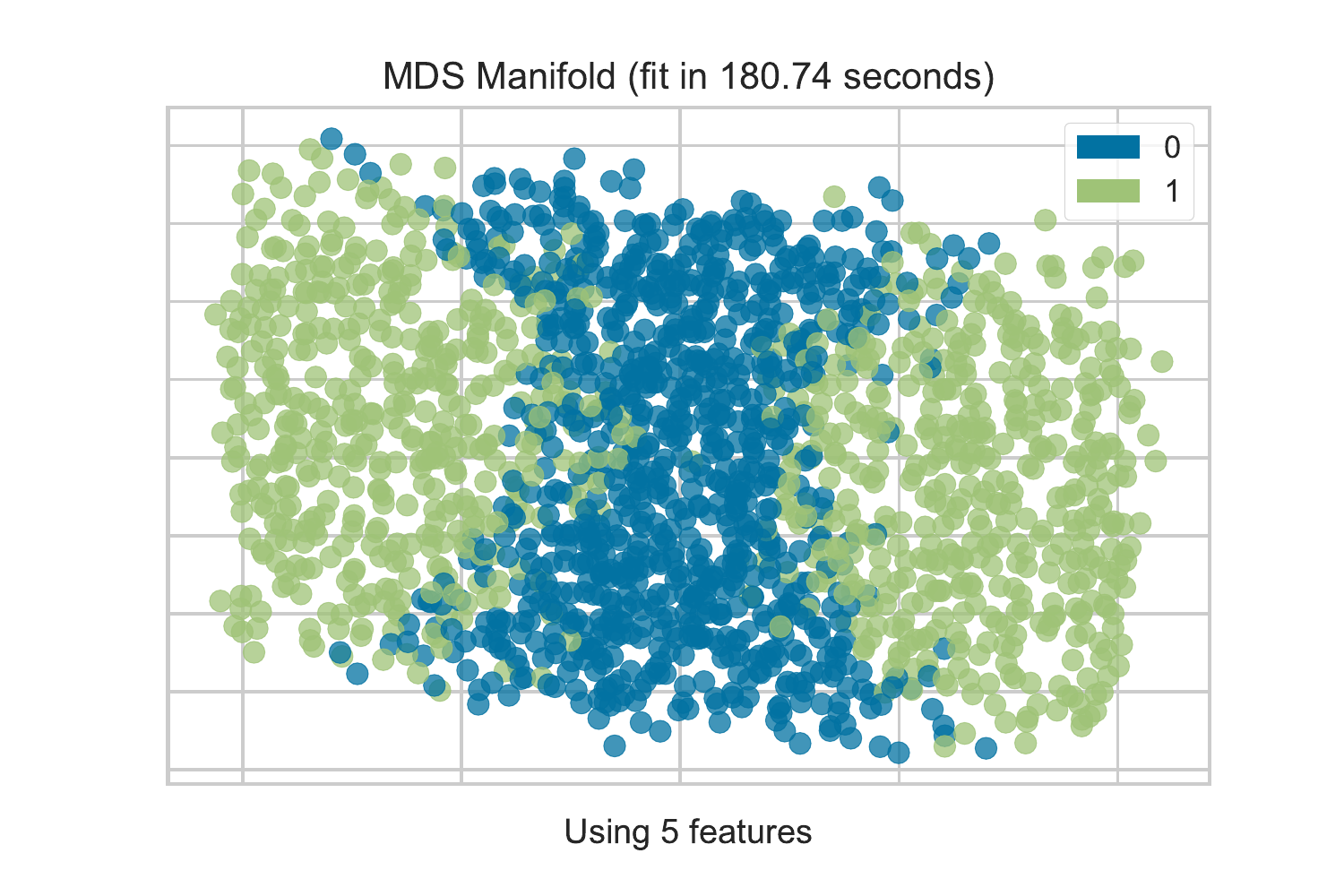}
    	\caption{}
    \end{subfigure}
    \begin{subfigure}{3.5cm}
    	\centering
    	\includegraphics[width=3.5cm]{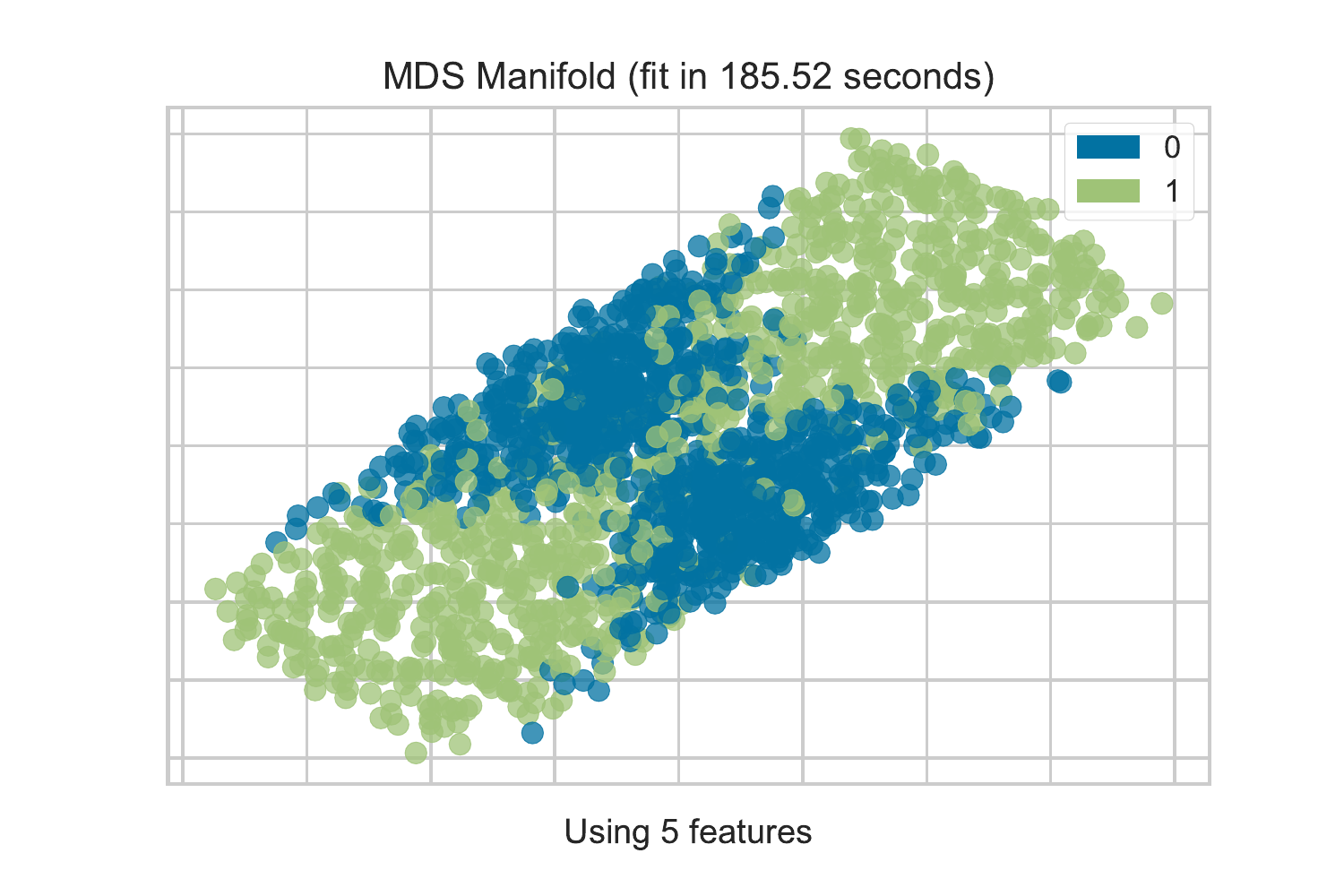}
    	\caption{}
    \end{subfigure}
    \begin{subfigure}{3.5cm}
    	\centering
    	\includegraphics[width=3.5cm]{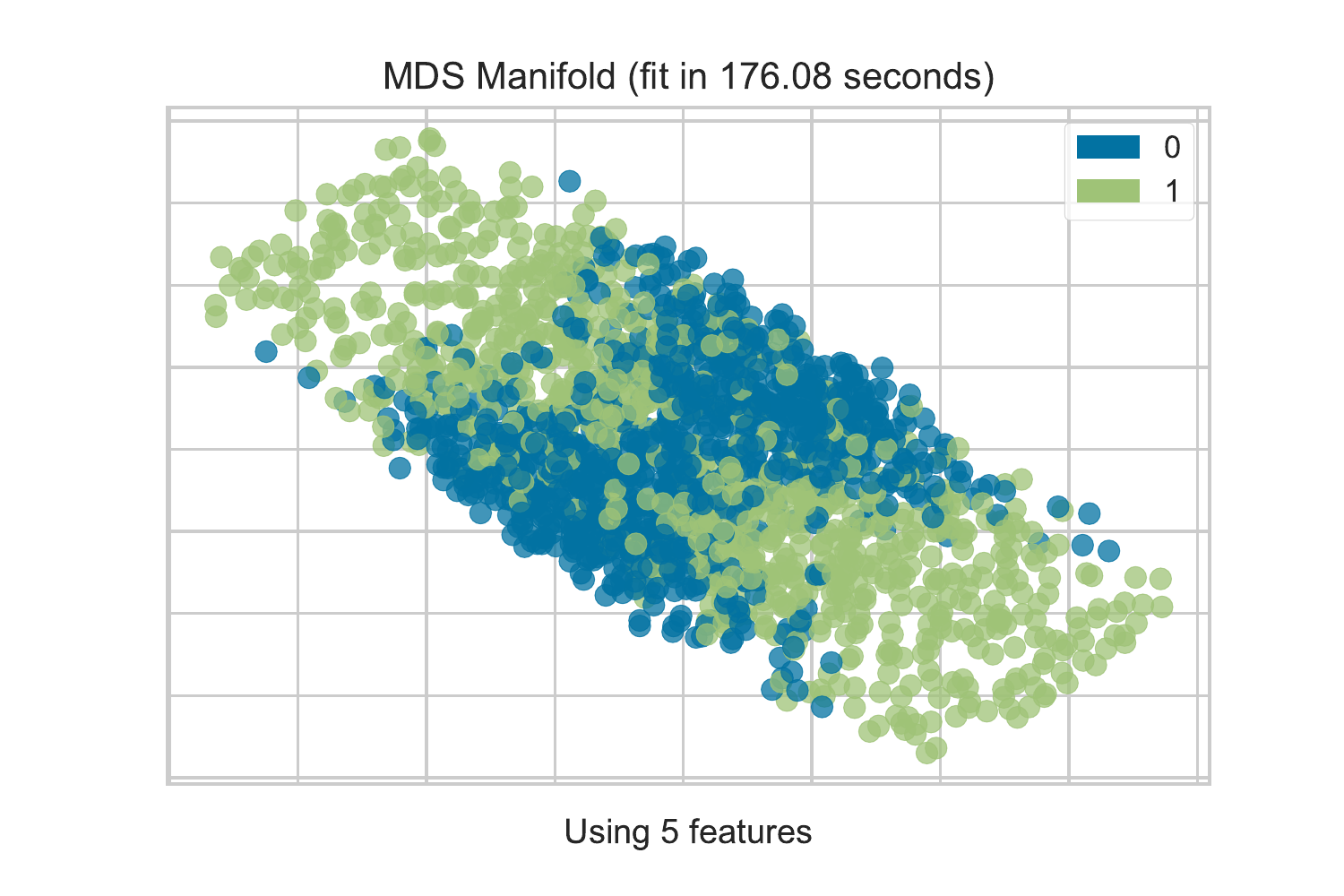}
    	\caption{}
    \end{subfigure}
    \caption{The MDS manifold projections for the datasets with (a) $n=2$, (b) $n=3$, (c) $n=4$ and (d) $n\rightarrow\infty$ (viz, the amoeba $\mathcal{A}_P$ itself).}\label{F0mdslargern}
\end{figure}

One may also try to apply such projection to higher $n$. As shown in Figure \ref{F0mdslargern}, the plots still have the similar scatterings to the one for $n=1$.

\subsection{Reproduction from Weights}\label{weights}
Since the MLP has a rather high accuracy predicting the genus, we may use the weight matrices and bias vectors to obtain a purely linear expression for the genus. Again, let us illustrate this with $n=1$. With the projected vector $(x,y)$ as input, it turns out that the minimal network with such performance only requires one hidden layer with four neurons. This means that after passing the data to the MLP, we get the expression
\begin{equation}
    p=W_2\cdot\text{ReLU}(W_1\cdot(x,y)^\text{T}+\bm{b}_1)+\bm{b}_2,
\end{equation}
where $W_{1,2}$ are weight matrices and $\bm{b}_{1,2}$ are the bias vectors. For instance, in our training, we obtain
\begin{equation}
    \begin{split}
        W_1&=\begin{pmatrix}
    0.4667&1.4986&-0.3313&-1.3031\\
    -1.3543&-1.0408&-0.5699&-1.5610
    \end{pmatrix}^\text{T},\\
    W_2&=\begin{pmatrix}
    -1.1199&0.6904&0.6216&0.2827
    \end{pmatrix}
    \end{split}
\end{equation}
and
\begin{equation}
    \bm{b}_1=(-0.0349,0.0439,0.3044,-0.5884)^\text{T},~\bm{b}_2=(-4.1003).
\end{equation}
Then the genus is simply given by $g=\theta(p)$ where $\theta$ is the Heaviside function.

One may also analyze the network structures for higher $n$. For example, still with fewer than 2000 training samples, an MLP (with a slightly larger size) can give over $0.9$ accuracy for $\mathcal{A}_P$, namely $n\rightarrow\infty$. The long expression determining the genus from the MLP was listed in \cite{Bao:2021olg}, and we shall not repeat it here.

Likewise, we can consider lattice polygons with more internal points which would lead to a larger range of the genus. The manifold learnings and the approximated expressions from the neural network can be applied to these cases in a similar manner although the analysis would be more involved. More examples can be found in \cite{Bao:2021olg}, along with CNN application to genus identification from Monte-Carlo generated images.

\section{Brain Webs for Brane Webs}\label{branewebs}

In Type IIB String Theory brane webs are configurations of $(p,q)$ 5-branes, containing 5-brane line segments and junctions, where three or more 5-branes meet. We assume that the $(p,q)$ 5-branes end on $(p,q)$ 7-branes which are depicted as dots in the plane. The $(p,q)$ charges of the 7-brane define an $SL(2,\mathbb{Z})$ monodromy action $M_{(p,q)}$ as one encircles it. We account for this by introducing a branch cut emanating from the 7-brane.

The data specifying a brane web can be encoded in a web matrix which lists the charges of the $(p,q)$ 7-branes, multiplied by the number $n$ of copies of each 5-brane.
\begin{equation}
\label{eq:web_matrix}
    W=\left(\begin{array}{cccc}
        n_1 p_1 & n_2 p_2 & \cdots & n_L p_L \\
        n_1 q_1 & n_2 q_2 & \cdots & n_L q_L
    \end{array}\right)\,,
\end{equation}
where $L$ is the number of external legs of the web\footnote{We work with the convention that all the charges of the 5-branes are ingoing and the legs are ordered anticlockwise around the junction.}.

A planar configuration of three or more 5-branes meeting at a point describes a 5d SCFT \cite{Aharony:1997ju,Aharony:1997bh}. Thus classifying 5-brane webs, corresponds to constructing 5d SCFTs. There are of course an infinite number of such webs but the space is hugely redundant, as infinitely many seemingly different $W$'s actually correspond to physically equivalent configurations. This is due to two reasons: firstly the SL$(2,\mathbb{Z})$ self-duality of Type IIB String Theory identifies congruent $W$'s; secondly 7-branes can be moved along the $(p,q)$ line past the intersection point in a Hanany-Witten move \cite{Hanany:1996ie} without changing the low energy physics. Thus the classification of webs requires to mod out the space of webs by these equivalence relations. 

The problem of classifying brane webs for $L=3$ is in fact equivalent to the classification of sets of 7-branes. In \cite{DeWolfe:1998eu} it was conjectured that inequivalent sets of 7-branes are characterised by the total monodromy of the webs, defined as the product of the monodromies of all the 7-branes 
\begin{equation}
    M_\text{tot} = M_{(p_{1},q_{1})} M_{(p_{2},q_{2})} \cdots M_{(p_{L},q_{L})}
\end{equation}
and the asymptotic charge invariant $\ell$, defined as 
\begin{equation}
    \ell=\gcd\left\lbrace \det\left(\begin{array}{cc}
        p_i & p_j \\
        q_i & q_j
    \end{array}\right)\,,\,\forall\, i,j \right\rbrace\,.
\end{equation}
However, as counter examples in \cite{Arias-Tamargo:2022qgb} explicitly show, these classifiers are not enough to fully specify a set of 7-branes, and are thus to be thought of as necessary but not sufficient conditions for equivalence. With this in mind, we define the following two notions of equivalent webs:
\begin{itemize}
    \item \textbf{Strong equivalence:} Two webs are strongly equivalent if they can be transformed into each other by means of any combination of SL$(2,\ZZ)$ and Hanany-Witten moves.
    
    \item \textbf{Weak equivalence:} Two webs are weakly equivalent if they have the same rank, $\ell$, and $M_\text{tot}$ up to SL$(2,\mathbb{Z})$.
\end{itemize}

The goal of \cite{Arias-Tamargo:2022qgb}, as reviewed here, was to teach a Siamese neural network (SNN) to identify equivalent webs, given by their web matrices \eqref{eq:web_matrix}, according to these two notions. Specifically an SNN was trained that took as input $2 \times 3$ matrices $W_{i}$ and produced a 10-dimensional embedding $\textbf{x}_{i} \in \mathbb{R}^{10}$ for each web matrix such that the Euclidean distance between embeddings of equivalent webs was minimised and the distanced between inequivalent webs maximised. Two webs were predicted to be equivalent if the squared Euclidean distance between their embeddings was less than some threshold value.

\subsection{Weak Equivalence}

Considering the simplest case of just two equal sized classes $\textbf{X}_{1}$ and $\textbf{X}_{2}$ of weakly equivalent webs (48 webs per class), the SNN was trained to determine whether a pair of webs from the set $\textbf{X}_{1} \cup \textbf{X}_{2}$ belonged to the same group. It was observed that, after training on 80\% of the data, the network determines weak equivalence with 100\% accuracy on the remaining 20\%. This result is visualised in Figure \ref{fig:TSNE_webs} by t-distributed stochastic neighbour embedding (t-SNE). The webs group together into two distinct weakly equivalent clusters.

Motivated by this result the investigation was extended to consider more classes, namely 14. For more than two equal sized classes there will be more inequivalent pairs of webs than equivalent and so to prevent an unbiased accuracy score the SNN was trained and tested on an equal number of equivalent and inequivalent pairs. The results dropped to a more modest 77\% but the network still performed significantly better than random guessing.

\subsection{Strong Equivalence}

\begin{figure}[h]
	\centering
    \begin{subfigure}{7cm}
    	\centering
    	\includegraphics[width=3.5cm]{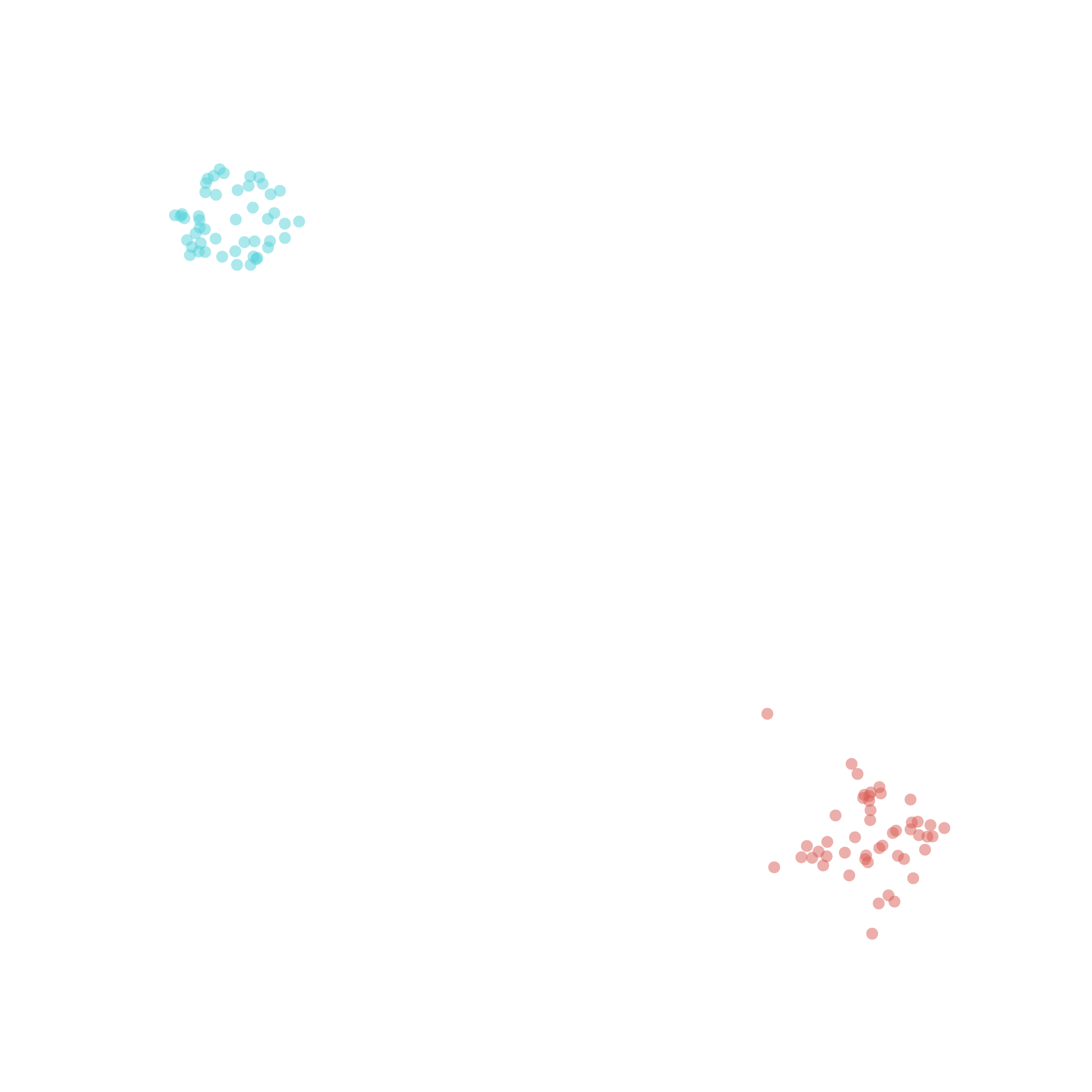}
    	\caption{$\textbf{X}_{1} \cup \textbf{X}_{2}$}
    \end{subfigure}
    \begin{subfigure}{7cm}
    	\centering
    	\includegraphics[width=3.5cm]{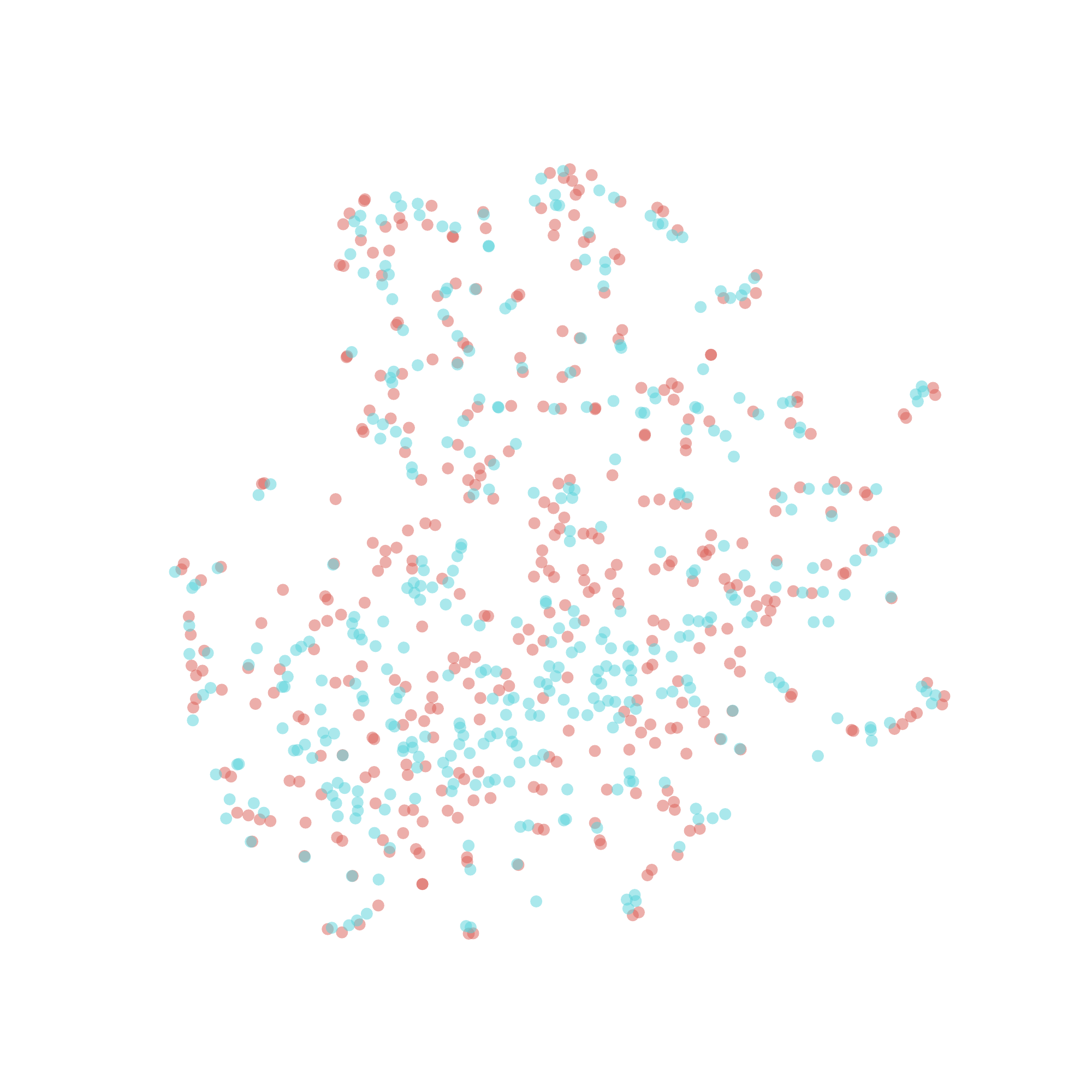}
    	\caption{$\textbf{Y}_{1} \cup \textbf{Y}_{2}$}
    \end{subfigure}
    \caption{t-SNE plots of the 10-dimensional web embeddings generated in the SNN of the webs from the respectively labelled datasets, reduced to 2 dimensions. }\label{fig:TSNE_webs}
\end{figure}

The same procedure was repeated using the notion of strong equivalence. With just two classes $\textbf{Y}_{1}$ and $\textbf{Y}_{2}$ (also 48 webs per class), the network predicted strong equivalence of pairs of webs from 20\% of $\textbf{Y}_{1} \cup \textbf{Y}_{2}$ with an accuracy of 50\%, after training on 80\%. This means that the network is no better than random guessing. It can be seen from the t-SNE plot in Figure \ref{fig:TSNE_webs} that the embeddings of $\textbf{Y}_{1}$ and $\textbf{Y}_{2}$ are mixed together and completely indistinguishable. Again the investigation was extended to 14 classes $\textbf{Y}_{I}, I=1,...,14$. Here the network predicted equivalence in the test set with accuracy of ~50\%.


\section{Quivers and Mutation}\label{quivermutation}

A rank $n$ cluster algebra is a constructively defined subalgebra of the field of rational functions in $n$ variables, with a set of generators (cluster variables) grouped into overlapping subsets (clusters) of size $n$, such that certain exchange relations allow one to transition from one cluster to another. These exchange relations, known as cluster mutations, can be described using the language of quivers.

Beyond representation theory, quivers physically provide a convenient representation of the particle content of gauge theories that describes D-branes probing varieties. A quiver diagram, $Q$, is a finite directed multigraph where each node represents a compact gauge factor U$(N_{i})$ of the total gauge group $\prod\limits_{i}{\text{U}(N_{i})}$ and edges represent bi-fundamental and adjoint chiral multiplets.

All quivers we consider have skew-symmetric exchange matrices and thus given an initial cluster $\{x_{1},…,x_{n}\}$, we can define the exchange relation replacing cluster variable $x_{j}$ by $x_{j}’$ in the cluster as
\begin{equation}
    x_{j} x_{j}’ = \prod_{i \rightarrow j \text{ in } Q}{x_{i}} + \prod_{j \rightarrow k \text{ in } Q}{x_{k}}
\end{equation}
for each $1 \leq j \leq n$, and where the products are over all incoming arrows and outgoing arrows, respectively. We thus get a new generator, $x_{j}’$, yielding a new cluster $\{x_{1},…,x_{j-1},x_{j}’,x_{j+1},…,x_{n}\}$. Note that this is not the most general form of cluster mutation, nor quiver mutation as one may include updates to the gauge group rank information. The process of mutation also updates the quiver by the following:
\begin{enumerate}
    \item Replace every incoming arrow $i \rightarrow j$ with the outgoing arrow $j \rightarrow i$ and replace every outgoing arrow $j \rightarrow k$ with the incoming arrow $k \rightarrow j$.
    \item For every 2-path $i \rightarrow j \rightarrow k$ add a new arrow $i \rightarrow k$.
    \item Remove all 2-cycles created by step 2.
\end{enumerate}

Given a quiver $Q$, we construct the associated cluster algebra $\mathcal{A}_{Q}$ by applying cluster mutation in all directions and iterating to obtain the full list of cluster variables. Generically, this process yields an infinite number of generators, i.e. cluster variables, as well as an infinite number of different quivers. Cluster algebras with a finite number of cluster variables are called finite type cluster algebras and cluster algebras of finite mutation type are those with a finite number of quivers. 

Cluster algebras and quiver gauge theories were brought together in \cite{Feng_2001,Feng_2001_2,Cachazo_2002}. It was shown that the cluster algebra mutation rules can be realised as Seiberg dualities of quiver gauge theories, where Seiberg duality is a generalisation of the classical electro-magnetic duality for 4d $\mathcal{N}=1$ supersymmetric gauge theories, this interpretation is the focus of the work in \cite{Bao:2020nbi} of which we summarise here. This work is also extended from quiver mutation to cluster mutation in \cite{Dechant:2022ccf}.

Our goal is to teach the machine to detect the mutation class of a given quiver. We consider two problems: the first is deciding whether two quivers are part of the same mutation class, and the second is determining specifically to which mutation class a given quiver belongs to. There are three types of quivers which we consider separately: finite type, finite mutation type which are not finite type (as finite mutation naturally include finite type), and infinite mutation type.

\subsection{Gathering Quiver Data}

The information of a quiver can be encoded in an adjacency matrix. It is these adjacency matrices together with the associated mutation class of the quiver that are fed to the machine. The list of quivers considered in the following investigations are shown in Figure \ref{fig:quivers} below. They are listed with an adjacency matrix representation and are labelled in the form $\textbf{Qi}$. 

\begin{figure}[h]
	\centering
	\includegraphics[width=7cm]{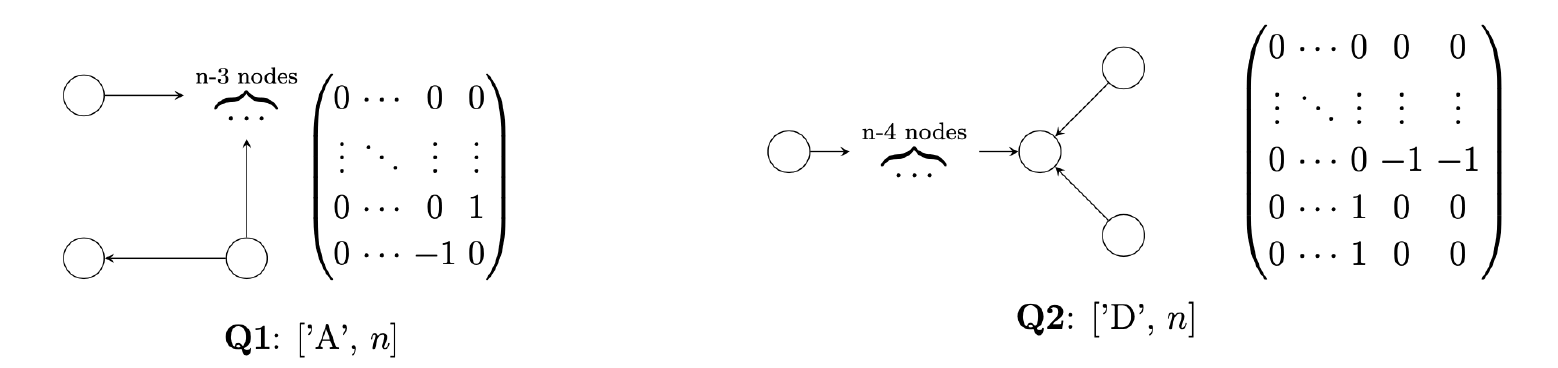}
	\includegraphics[width=7cm]{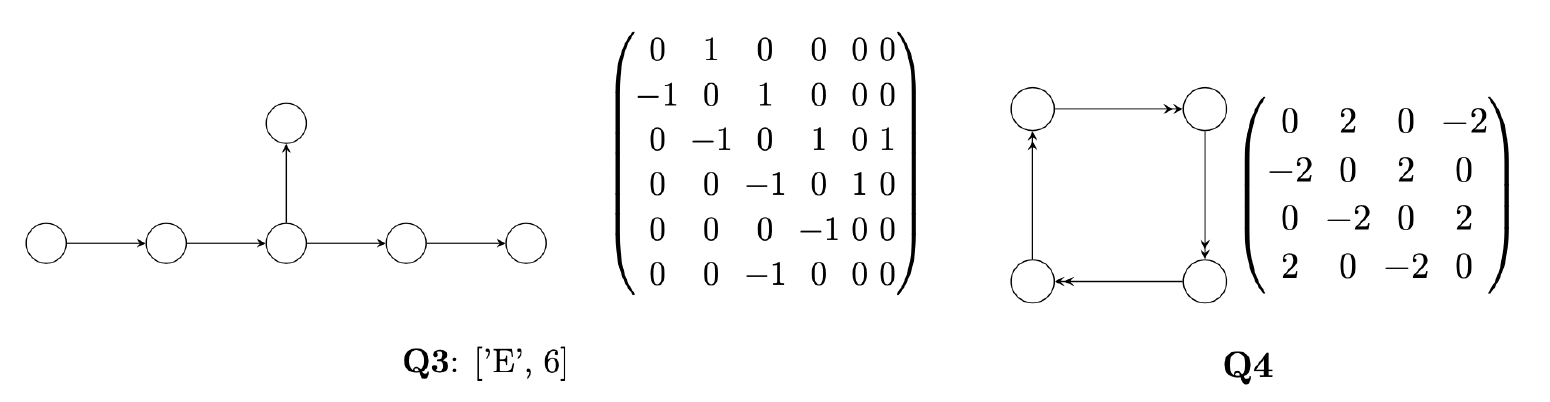}
	\includegraphics[width=7cm]{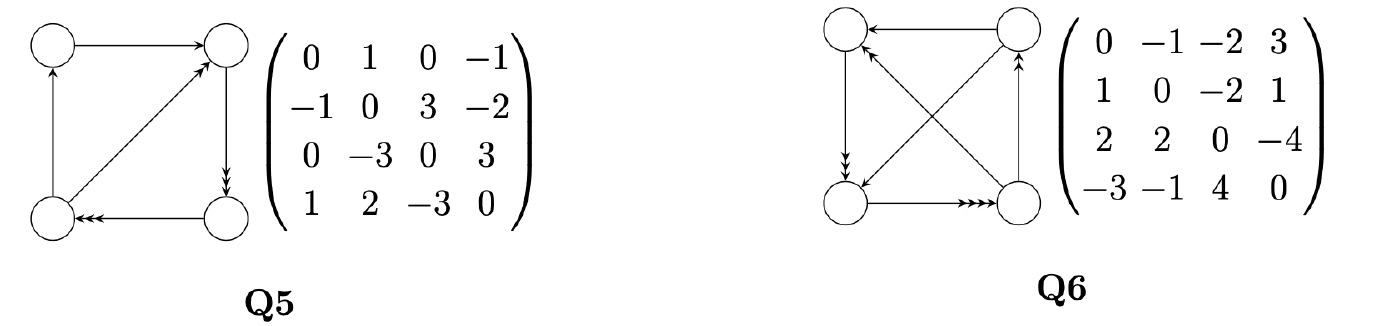}
	\includegraphics[width=5cm]{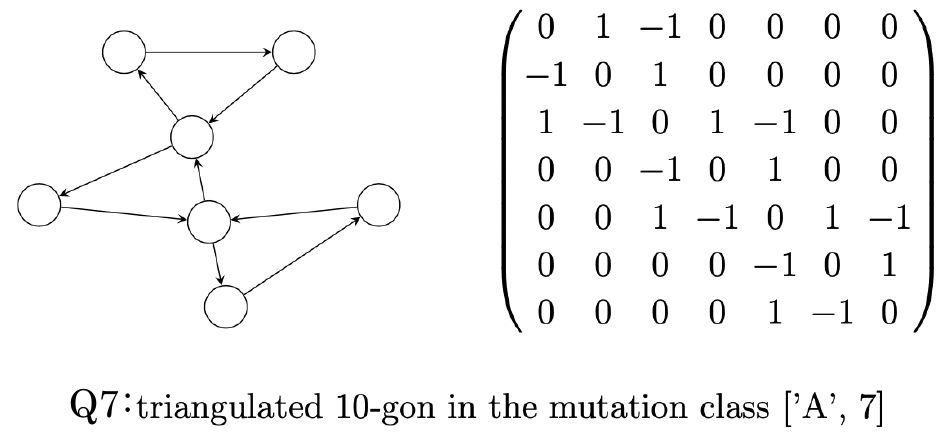}
	\includegraphics[width=5cm]{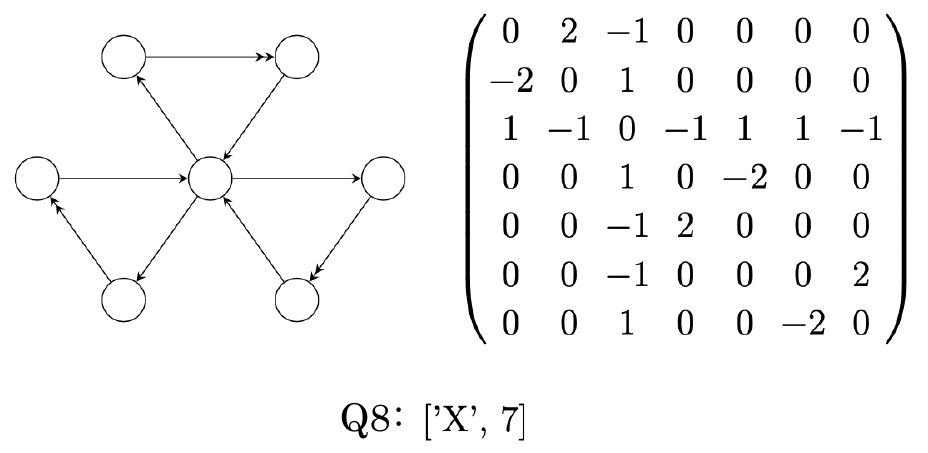}
	\includegraphics[width=5cm]{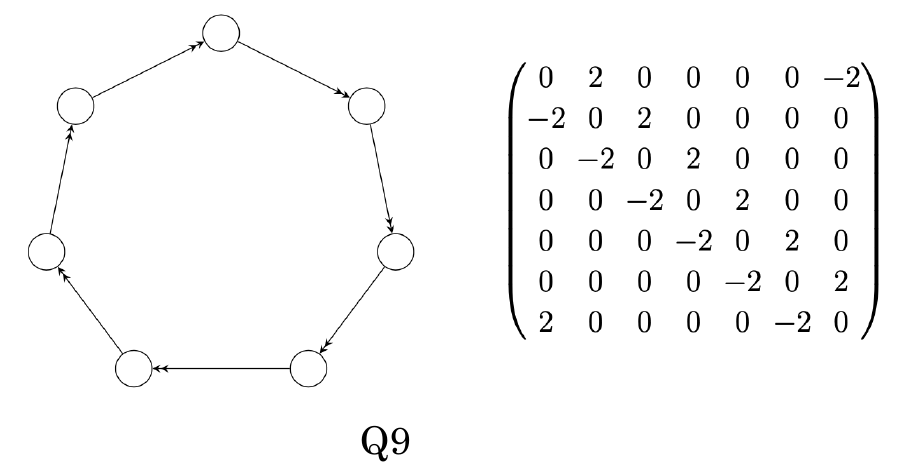}
    \caption{Quivers}\label{fig:quivers}
\end{figure}

\subsection{Classifying Mutation Classes}

We start off by considering the binary classification task where the input data consists of pairs of adjacency matrices $M_{1},M_{2}$ that describe two quivers and the output is either 0 or 1, where 1 indicates that $M_{1}$ and $M_{2}$ are in the same Seiberg duality class and 0 indicates that they are not. We apply the \texttt{Mathematica} built-in function \texttt{Classify} with Naïve Bayes (NB) method specified to carry out this classification.

On the two finite mutation type classes A4 and D4 ($\textbf{Q1}$ and $\textbf{Q2}$ in Figure \ref{fig:quivers}), this gives 100\% accuracy over 5-fold cross validation. For the case of two infinite type mutation classes, namely $\textbf{Q4}$ and $\textbf{Q5}$ in Figure \ref{fig:quivers}, we also get 100\% accuracy.

So far we have looked at one type at a time. We wonder whether different types would affect our results. A simple check would involve only two mutation classes with one Dynkin (i.e. finite type) and one affine (i.e. finite-mutation type but not finite). For instance, we test D4 and A(3,1)1 here. The \texttt{Classify} function again gives 100\% accuracy as for the example of A4 and D4. This is an exciting result. We find that learning mutation classes of the same type (e.g. only Dynkin) and learning those of different types (e.g. Dynkin + affine) have the same performance. Let us further try an example with one finite type D4 and one infinite mutation type, namely $\textbf{Q4}$. We still get perfect accuracy. It appears that the mutation types do not really affect our learning performance for NB.

We now contemplate datasets containing more mutation classes. For the finite type we use A6, D6 and E6, corresponding to $\textbf{Q1}, \textbf{Q2}$, and $\textbf{Q3}$ respectively. The learning result of 5-fold validation is 90\% which, albeit is not as perfect as the case with two classes, is still very good. For the infinite mutation type we add $\textbf{Q6}$ to $\textbf{Q4}$ and $\textbf{Q5}$ and we also see a drop from 100\% to 90\% accuracy. We see that the numbers of different classes does affect the performance of NB in determining whether two quivers are of the same mutation class. 

\subsection{Multiclass Classification}

Besides pairing matrices and assigning 1's and 0's, there is a more direct way to classify theories, we can simply assign different mutation classes with different labels and then let the machine tell which classes the given quivers belong to. We turn to \texttt{Python} to perform this task using a CNN with the help of \texttt{Sage} \cite{sage,musiker2011compendium} and \texttt{Tensorflow}. 

We choose three classes, namely $\textbf{Q7},\textbf{Q8}$ and $\textbf{Q9}$ in Figure \ref{fig:quivers}, the first two classes are finite while the third is infinite. We label the three classes $[1,0,0], [0,1,0]$ and $[0,0,1]$ respectively. Thus when the machine predicts $[a_{1},a_{2},a_{3}]$, it is giving probabilities of the three classes respectively. We find that there is only $\sim 55\%$ accuracy when the machine is trained on 80\% of the data. However, remarkably on the last class, which is infinite, the machine has 100\% accuracy, i.e. it always correctly recognises the matrices in this class and never misclassifies other matrices to this class. Hence the machine seems to have learnt something related to finite and infinite mutations.

\section{Dessins d'Enfants}\label{dessins}
Dessins d'Enfants, or children's drawings, are bipartite graphs drawn on Riemann surfaces that graphically represent the degeneracy information of Bely\v{\i} maps from the surface to the Riemann sphere. Where these Bely\v{\i} maps exist, via Bely\v{\i}'s theorem, one can define the Riemann surface using exclusively algebraic numbers $\overline{\mathbb{Q}}$ \cite{belyi}.

These dessins were hypothesised by Grothendieck to faithfully represent the absolute Galois group over the rationals, a famously elusive object \cite{esquisse}. Although the dessins are easy to draw, the computation of the underlying Bely\v{\i} map is notoriously difficult; and hence extracting any information about this map from the dessin alone is especially advantageous for uncovering the mysteries behind their connections to Galois groups.

It should be noted that dessins have an array of applications in physics also. They may be considered an interpretation of dimer models \cite{Franco:2005sm,Hanany:2011ra,Jejjala:2010vb,Bao:2021gxf}, arise naturally when working with K3 surfaces \cite{He:2012kw,He:2012jn} which may act as the moduli spaces for specific quiver gauge theories, and have been useful in factorising Seiberg-Witten curves \cite{Ashok:2006br,Bao:2021vxt}.

In this section we review the work of \cite{He:2020eva}, which seeks to apply supervised machine learning techniques to this programme of recovering Bely\v{\i} map information from the respective dessins. The focus is on extracting the size of the Galois orbits that the dessins belong to from tensor representations of the graphs. The dataset of dessins used arises from elliptic K3 surfaces, where surprisingly their $j$-invariants are Bely\v{\i} in nature and hence produce dessins.

\subsection{Defining the Data}
Bely\v{\i}'s theorem states that a Riemann surface, $X$, can be defined over algebraic numbers if and only if there exists a map to the Riemann sphere, $\beta: X \longmapsto \mathbb{P}^1_{\mathbb{C}}$, which has exactly three ramification points. Due to the symmetry under the M\"obius group of the Riemann sphere, these three ramification points can be mapped to $\{0,1,\infty\}$. Then the preimages for the interval [0,1] on the Riemann sphere through the Bely\v{\i} map, $\beta$, can be calculated. 

Each of these preimages on the Riemann sphere can be affiliated to the bipartite structure of the dessin:
$\beta^{-1}(0) \mapsto \circ \,,\quad \beta^{-1}(1) \mapsto \bullet \,,\quad \beta^{-1}(0,1) \mapsto - \ $. Within this structure a vertex's degree indicates the point's ramification index, i.e. the Taylor expansion of $\beta$ about that preimage point beyond the constant term will begin at some order $n>1$, being the ramification index. Further to this each face is associated to a preimage of $\infty$, and half the number of bounding edges to the face indicates that point's ramification index.

\begin{figure}[tb]
	\centering
	\begin{subfigure}{0.45\textwidth}
    	\centering
    	\includegraphics[width=0.5\textwidth]{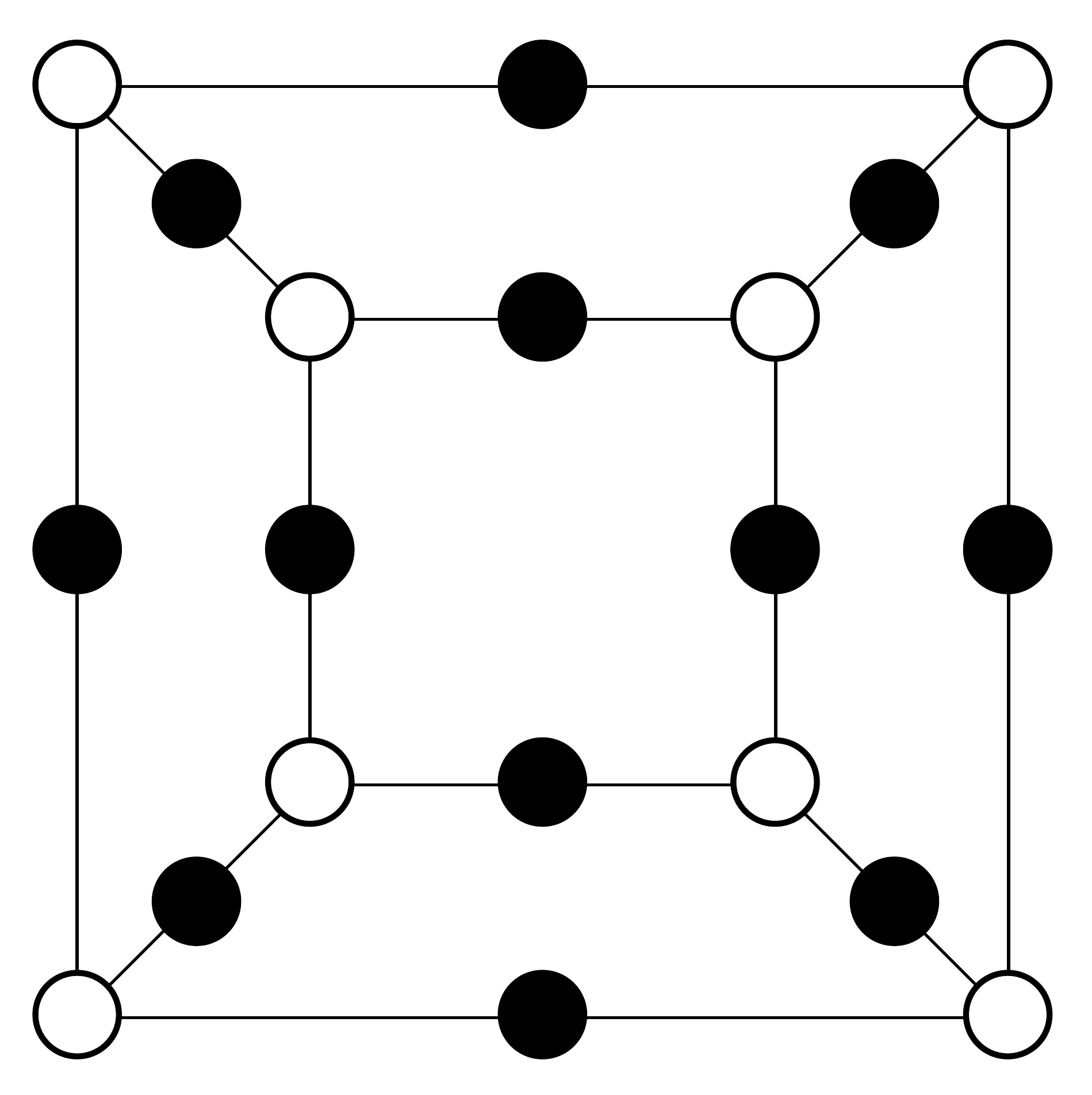}
    	\caption{}\label{fig:dessin}
	\end{subfigure} 
    \begin{subfigure}{0.45\textwidth}
    	\centering
    	\includegraphics[width=0.5\textwidth]{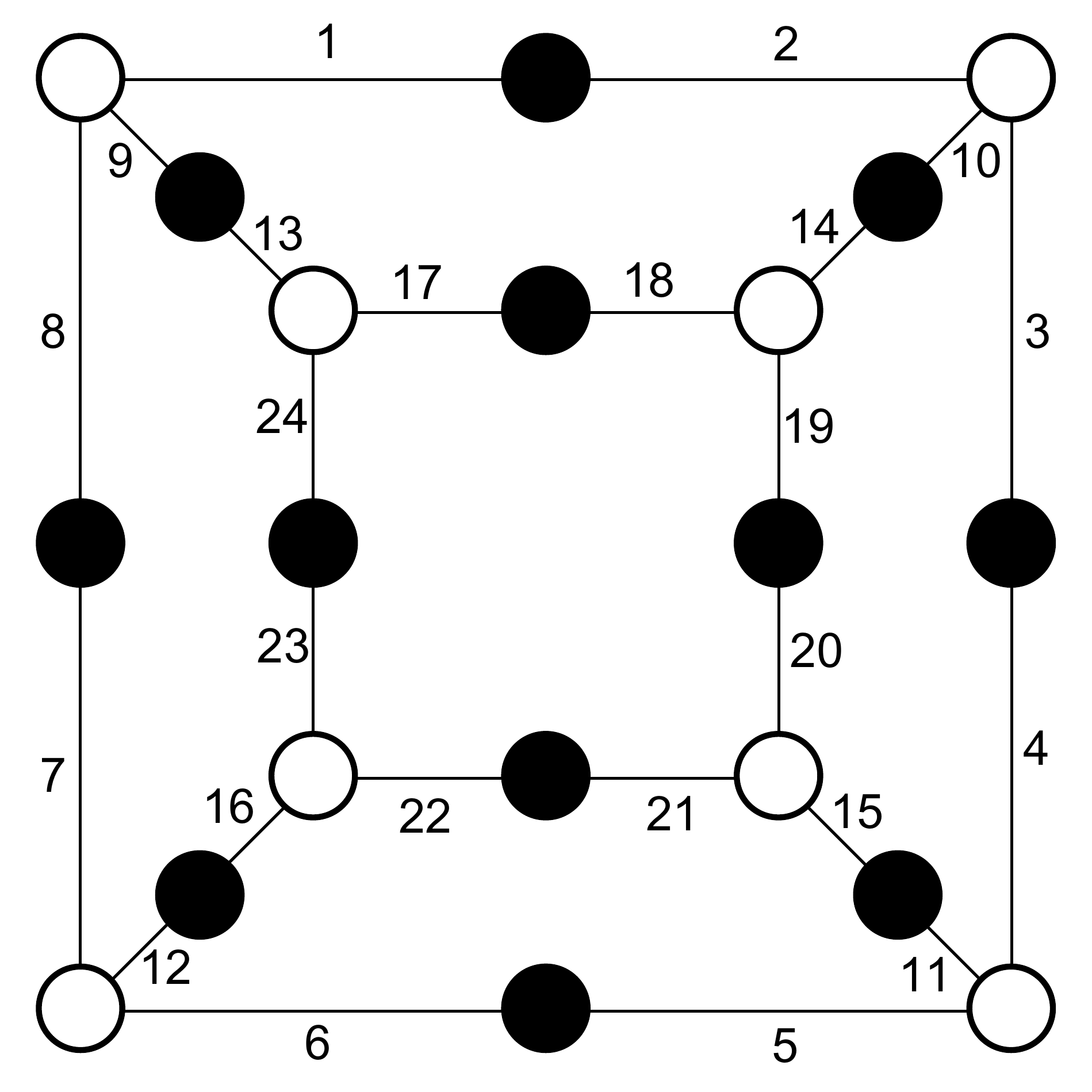}
    	\caption{}\label{fig:dessin_cyclic}
    \end{subfigure}
\caption{An example dessin d'enfant (a), of the type examined in this work, it is a planar dessin, and also clean - as all same colour nodes have the same valency. Then the same example dessin with edges labelled (b) for the cyclic edge list representation:
\{\{\{8,1,9\},\{2,3,10\},\{4,5,11\},\{12,6,7\},\{18,14,19\},\{22,16,23\}, \\
\{13,24,17\},\{21,20,15\}\}, \{\{1,2\},\{9,13\},\{10,14\},\{3,4\},\{15,11\},\{24,23\},\{17,18\},\{8,7\},\{5,6\},\{12,16\},\{21,22\}, \\
\{20,19\}\}\}. }\label{fig:dessins}
\end{figure}

As an example consider the Bely\v{\i} map
\begin{equation}\label{betaeg}
    \beta(z) = \frac{(z^8-14z^4+1)^3}{(-108z^4(z^4+1)^4)}\,,
\end{equation}
which has ramification information $\{3^8|2^{12}|4^6\}$ and is drawn in Figure \ref{fig:dessin}. The ramification information indicates the shown 8 white vertices of valence 3, 12 black vertices of valence 2, and 6 faces  with 8 bounding edges (including the external face as truly drawn on the Riemann sphere).

As can be seen in this example, $\beta$ is defined over $\mathbb{Q}$. However in other examples the map may require non-rational numbers. Each of these non-rational numbers in the map definition amount to a field extension of $\mathbb{Q}$ towards $\overline{\mathbb{Q}}$. The minimal polynomial over $\mathbb{Q}$ with each non-rational number as a root will have other roots also, and changing the root occurring in the map to all the other roots amounts to taking this Bely\v{\i} map, and respective dessin which changes each time, through its Galois orbit. For the example above as no non-rational numbers occur the Galois orbit is size 1, and in the database considered they vary up to size 4.

The dataset machine learnt comes from \cite{He:2012jn}. These 191 dessins come from the modular group PSL$(2,\mathbb{Z})$, specifically each of the 112 index 24 torsion-free genus-zero subgroups. Performing quotient action by each of these subgroups on the upper half-plane fibred with a complex line produces an elliptic K3 surface whose $j$-invariant is surprisingly always a Bely\v{\i} map! Then taking each of these maps through their Galois orbits produces the 191 dessins considered.

\subsection{Learning Galois Orbit Size}
These 191 dessins each have a respective Galois orbit size from $\{1,2,3,4\}$. The machine learning approach was hence designed to apply a supervised neural network classifier to learning this Galois orbit size from an input dessin represented tensorially.

The initial method of representation ignored the bivalent black nodes, absorbing them into the edges, and represented each dessin by its adjacency matrix on the white nodes, $\mathcal{M}_{\text{8x8}}$.
However, even with data augmentation to increase the data size, the learning was not strong with performance only marginally better than random guessing. This poor performance was attributed to the adjacency matrix representation not being faithful, i.e. there were occasional scenarios where different dessins had the same adjacency matrix. 

To design a faithful representation method, the category equivalence between dessins and finite sets under permutation was used. This time all the edges were labelled, and at each white node the edges were listed clockwise, and at each black node anticlockwise (although direction redundant here as they are bivalent). This representation method is shown in Figure \ref{fig:dessin_cyclic}, and admitted its own methods for data augmentation of relabelling edges, cyclically shuffling the edge lists, or shuffling the node lists. The edge lists were then flattened to give the final input vectors passed as input to the NNs, $\mathcal{V}_{48}$. 

For the adjacency matrix format, augmentation of the 152 distinct dessin matrices with $\sim 1000$ permutations of rows/columns (excluding those which didn't change the matrix) lead to a proportional distribution over the orbit size classes of: $[0.49,0.31,0.19,0.01]$. Conversely, the equivalent data distribution after augmenting the edge list representation by 1000 permutations was: $[0.40,0.35,0.23,0.02]$.

The NN classifiers of 4 layers of 512 neurons with LeakyReLU activation and 0.25 dropout were trained using an Adam optimiser over the cross-entropy loss. Due to the bias of the data sets towards lower orbit size classes, as well as the prototypical accuracy metric MCC was also used and provides a better measure of the learning performance. In addition 5-fold cross-validation was performed for each investigation such that the measures could be averaged and standard error calculated to provide confidence in the results.

Whilst the adjacency matrix results were only marginally better than random guessing (exemplified by an MCC close to 0), the results using the cyclic edge list representation were far better, as shown in Table \ref{table:dessin_results}. The performance indicates that the Galois orbit size is perhaps combinatorically encoded in the dessin information, in such a way that this simple ML architecture can take advantage of it.

\begin{table}[tb]
\centering
\begin{tabular}{|c|cc|}
\hline
\multirow{2}{*}{\begin{tabular}[c]{@{}c@{}}Data\\ Format\end{tabular}} & \multicolumn{2}{c|}{Performance Measure} \\ \cline{2-3} & \multicolumn{1}{c|}{Accuracy}  & MCC \\ \hline
$\mathcal{M_{\text{8x8}}}$  & \multicolumn{1}{c|}{\begin{tabular}[c]{@{}c@{}} $0.536$ \\ $\pm 0.002$ \end{tabular}} & \begin{tabular}[c]{@{}c@{}} $0.180$ \\ $\pm 0.008$ \end{tabular} \\ \hline
$\mathcal{V}_{48}$  & \multicolumn{1}{c|}{\begin{tabular}[c]{@{}c@{}} $0.920$ \\ $\pm 0.003$ \end{tabular}} & \begin{tabular}[c]{@{}c@{}} $0.880$ \\ $\pm 0.040$ \end{tabular} \\ \hline
\end{tabular}
\caption{Accuracy and MCC learning results for the investigations, constituting 2 primary data formats: the adjacency matrix format $\mathcal{M_{\text{8x8}}}$ and the cyclic edge list format $\mathcal{V}_{48}$.
}\label{table:dessin_results}
\end{table}

\section{Hessian Manifolds, Optimal Transport and GANs}\label{hessiangan}
In \cite{2016arXiv160405645H,2016arXiv160702923H,2016arXiv160807209H,2017arXiv171109881H}, the optimal transport problem and Monge-Amp\`ere equations were related to K\"ahler geometry. In particular, consider a compact toric manifold $\mathcal{M}$ with K\"ahler metric invariant under $T^n$ in the $(\mathbb{C}^*)^n$-action. Then the study of this complex manifold could mostly be reduced to the convex analysis on $\mathbb{R}^n$. A plurisubharmonic function $\psi$ invariant under $T^n$ can be represented by some convex function $\phi$ on $\mathbb{R}^n$ via\footnote{There could be different conventions for the constant $c_n$, such as $c_n=\frac{i^n}{(2\pi)^nn!}$ in \cite{2013arXiv1302.4045B}. However, this is not really important here as it could be absorbed into $\psi$ or $\phi$.} \cite{2013arXiv1302.4045B}
\begin{equation}
	\text{Log}_*c_n(\partial\bar{\partial}\psi)^n=\det\left(\frac{\partial^2\phi}{\partial x_i\partial x_j}\right)\text{d}V,
\end{equation}
where $x_i$'s are the coordinates on $\mathbb{R}^n$ and $\text{d}V$ denotes the Euclidean volume form on $\mathbb{C}^n$. Here, the map
\begin{equation}
	\text{Log}:~(\mathbb{C}^*)^n\rightarrow\mathbb{R}^n,~z\mapsto\log(z)=:x
\end{equation}
is defined such that the $j^\text{th}$ component of $x$ satisfies $x^j=\log\left(|z^j|^2\right)$. By an abuse of notation, we abbreviate the above equation as
\begin{equation}
	c_n(\partial\bar{\partial}\psi)^n=\det\left(\frac{\partial^2\phi}{\partial x_i\partial x_j}\right)\text{d}V.\label{comp2real}
\end{equation}
Therefore, the \emph{complex} Monge-Amp\`ere operator in this case can be represented by the Hessian determinant of the real $\phi$, or in other words, reduced to the \emph{real} Monge-Amp\`ere equation
\begin{equation}
	\det\left(\frac{\partial^2\phi}{\partial x_i\partial x_j}\right)=\text{e}^{-\lambda\phi}
\end{equation}
for $\lambda\in\mathbb{R}$. In fact, this endows $\mathcal{M}$ with a Hessian metric which reads
\begin{equation}
	g=\nabla\text{d}\phi_k=\frac{\partial^2\phi_k}{\partial x_i\partial x_j}\text{d}x_i\otimes\text{d}x_j
\end{equation}
for some smooth $\phi_k:U_k\rightarrow\mathbb{R}$ in terms of the local covering $\{U_k\}$ in the charts. The right hand side of \eqref{comp2real} is known as the real Monge-Amp\`ere measure MA$(\phi)$ of the Hessian metric, where $\phi$ is the shorthand notation for $\{\phi_k\}$. Moreover, we require $\mathcal{M}$ to be special\footnote{An affine manifold is said to be special if d$V$ is preserved by the transition maps, viz, Hol$(\nabla)\subset\text{SL}(n,\mathbb{R})$.} such that MA$(\phi)$ is invariant under coordinate transformations and hence globally defined on $\mathcal{M}$. The existence and uniqueness of such MA$(\phi)$ was proven in \cite{Cheng1980}.

\subsection{Optimal Transport}\label{othess}
The aim of the optimal transport problem is to minimize some cost function $c:X\times Y\rightarrow\mathbb{R}$, for some probability spaces $X$ and $Y$. In the setting of Hessian manifolds, $X=\mathcal{M}$ and $Y=\mathcal{M}^*$, where $\mathcal{M}^*$ is the dual Hessian manifold \cite{2016arXiv160702923H}. We shall not expound the details of dual Hessian manifolds here, and it suffices to know that $\text{d}\phi:\mathcal{M}\rightarrow\mathcal{M}^*$ is a diffeomorphism. Given the Borel probability measures $\mu$ and $\nu$ on $\mathcal{M}$ and $\mathcal{M}^*$ respectively, we can define
\begin{definition}
	The \emph{$\nu$-Monge-Amp\`ere measure} of a convex section $\phi$ is $\textup{MA}_\nu(\phi)=(T_\phi)_*\nu$. Here $T_\phi$, which can be identified as $\textup{d}(\phi^*)$, is the inverse of $\textup{d}\phi$ (defined almost everywhere on $\mathcal{M}^*$).
\end{definition}
We now state the optimal transport problem and the associated cost function.
\begin{definition}
	Let $(\mathcal{M},L)$ be a compact Hessian manifold with convex section $\phi_0$ and $\Pi$ the fundamental group of $\mathcal{M}$. The cost function $c:\mathcal{M}\times\mathcal{M}^*\rightarrow\mathbb{R}$ is $c(x,y)=-[x,y]+\phi_0(x)+\phi^*_0(y)$. The pairing operation is defined as $[x,y]:=\sup\limits_{\gamma\in\Pi}\gamma\cdot q(x)-q$ for $q$ being a point in the fibre of $K^*$ over $y$, and $\phi_0^*(y)=[x,y]-\phi_0(x)$ is the Legendre transform\footnote{The definition of [-,-] is independent of the choice of $q$. Hence this is well-defined.}. Then the optimal transport problem is to minimize the transportation cost $I_c=\int_{\mathcal{M}\times\mathcal{M}^*}c(x,y)\textup{d}\rho$ for all probability meausures $\rho$ on $\mathcal{M}\times\mathcal{M}^*$.
\end{definition}

Let $\phi$ be in the K\"ahler class of $\phi_0$. Consider the cost function $c'(x,y)$ induced by $\phi$. Then $c'(x,y)=c(x,y)-f-f^c$, where $f:=(\phi-\phi_0)$ and $f^c:=(\phi^*-\phi^*_0)$. In terms of the Kantorovich problem in optimal transport, this means that $c'$ is equivalent to $c$. In other words, $c'$ minimizes $I_c$ as well since
\begin{equation}
	\begin{split}
	    I_{c'}&=\int_{\mathcal{M}\times\mathcal{M}^*}c'(x,y)\textup{d}\rho=\int_{\mathcal{M}\times\mathcal{M}^*}c(x,y)\textup{d}\rho-\int_{\mathcal{M}}f\textup{d}\rho-\int_{\mathcal{M}^*}f^c\textup{d}\rho\\
	&=\int_{\mathcal{M}\times\mathcal{M}^*}c(x,y)\textup{d}\rho-\int_{\mathcal{M}}f\textup{d}\mu-\int_{\mathcal{M}^*}f^c\textup{d}\nu=I_c+C,
	\end{split}
\end{equation}
where $C$ is a constant independent of $\rho$.

It was shown in \cite{2016arXiv160702923H} that the cost function is the squared distance induced by the flat Riemannian metric $d$ determined by $L$, viz, $c(x,y)=\frac{d^2(x,y)}{2}$. From above, we can also write the dual Kantorovich problem, that is, to maximize the quantity $J=\int_{\mathcal{M}}f\text{d}\mu+\int_{\mathcal{M}^*}f^c\text{d}\nu$, where $f$ is known as the Kantorovich potential and $f^c$ is called its $c$-transform.

Now we can relate the Monge-Amp\`ere measures to the optimal transport problem \cite{2016arXiv160702923H}.
\begin{theorem}
	There exists a smooth, strictly convex section $\phi$ such that $\textup{MA}_\nu(\phi)=(T_\phi)_*\nu=\mu$. Then $\textup{d}\phi:\mathcal{M}\rightarrow\mathcal{M}^*$ is the optimal transport map determined by $\mu$, $\nu$ and $c(x,y)$. In other words, such $\phi$ minimizes the transportation cost $I_c$.
\end{theorem}

Here, we discussed how special Hessian manifolds can be studied via optimal transport. It would also be interesting to study the relations of real Monge-Amp\`ere equations on real tori and complex Monge-Amp\`ere equations on Abelian manifolds as in \cite{2016arXiv160405645H}. In terms of mirror symmetry and tropical geometry, solutions to real Monge-Amp\`ere equations on Hessian manifolds appear as large complex limits of K\"ahler-Einstein metrics on complex mainfolds \cite{aspinwall2009dirichlet}. It would also be interesting to extend the study to singular or non-compact settings in the context of mirror symmetry.

\subsection{GANs}\label{gan}
A generative adversarial network (GAN) \cite{goodfellow2014generative} is composed of two neural networks known as the generator and the discriminator. Given the training data, the generator captures its distribution from the latent space and generates new samples. On the other hand, the discriminator estimates the probability of the samples coming from the true data or from the generator. The two networks are trained simultaneously, and the competition between them improves the performance of the machine until the generated samples cannot be distinguished from the true data.

In \cite{DBLP:journals/corr/abs-1710-05488}, the improvement of the GAN model was attempted using the theory of optimal transport. Let $X$ be the image space as a manifold embedded in $\mathbb{R}^n$. The latent space is then some $\mathbb{R}^m$ which is related to $X$ under the usual homeomorphism in the charts $\psi_i:U_i\rightarrow\mathbb{R}^m$, and hence $\psi_i$ corresponds to the decoder NN. Then the generative network and the discriminative network both have certain data distributions, say, $\mu$ and $\nu$. The goal of GANs is therefore to minimize the Wasserstein cost $W_c=\int_{X\times X}c(x,y)\text{d}\rho$ for some cost function $c:X\times X\rightarrow\mathbb{R}$, where $\rho$ is the Borel probability measures on $X\times X$. Now that this becomes an optimal transport problem, it is natural to have a correspondence between the geometry side and NN side.

\subsection{The Dictionary}\label{dictionary}
Since $\mathcal{M}$ and $\mathcal{M}^*$ are equivalent as affine manifolds, both of them correspond to the image space $X$ inside the ambient space. It is also natural to match the corresponding probability measures $\mu_\text{hess}\leftrightarrow\mu_\text{NN}$ and $\nu_\text{hess}\leftrightarrow\nu_\text{NN}$.

We can then equate the transportation costs $I_c\leftrightarrow W_c$ as well. In other words, the two cost functions should also match. Indeed, in the language of \cite{DBLP:journals/corr/abs-1710-05488}, we have $c_\text{NN}(\bm{x},\bm{y})=|\bm{x}-\bm{y}|^2/2$. As argued in \cite{DBLP:journals/corr/abs-1710-05488}, the discriminator function, denoted by $\phi_\xi$, which is the metric between distributions in the discriminator, is equivalent to the Kantorovich potential. Hence, for the dual Kantorovich problem, we have $f\leftrightarrow\phi_\xi$ as well as $f^c=c_\text{hess}(x,y)-f\leftrightarrow\phi_\xi^c=c_\text{NN}(x,y)-\phi_\xi$.

In light of \cite{DBLP:journals/corr/abs-1710-05488}, the correpsonding Monge-Amp\`ere equation for GAN is
\begin{equation}
	\det\left(\frac{\partial^2u}{\partial x_i\partial x_j}\right)=\frac{\mu(x)}{\nu(\nabla u(x))},
\end{equation}
where $u$ is called the Brenier potential, and $\nabla u$ is the transportation map that optimizes our problem. In the GAN model, $u$ is often represented by linear combinations and ReLUs. Then we can write the correspondence for the two Brenier problems via $\phi\leftrightarrow u$ and $\text{d}\phi\leftrightarrow\nabla u$. Recall that $T_\phi$ is the inverse of $\text{d}\phi$. Thus, the Brenier problem formulated by $(\nabla u)_\sharp\mu=\nu$ in \cite{DBLP:journals/corr/abs-1710-05488} should correpsond to $(T_\phi)_*\nu=\mu$ in the dictionary. Indeed, solving these two equations gives the optimal transport maps of the two problems.

Since the K\"ahler potential, say $\psi$, is plush, in terms of \eqref{comp2real}, we could write
\begin{equation}
	c_n(\partial\bar{\partial}\psi)^n=\frac{\mu(x)}{\nu(\nabla u(x))}\text{d}V,
\end{equation}
whose left hand side could be expanded as
\begin{equation}
	\text{LHS}=c_n\epsilon^{\mu_1\dots\mu_n\nu_1\dots\nu_n}\prod_{i,j}\frac{\partial^2\psi}{\partial z^{\mu_i}\partial\bar{z}^{\nu_j}}\text{d}z^1\wedge\dots\wedge\text{d}z^n\wedge\text{d}\bar{z}^1\wedge\dots\wedge\text{d}\bar{z}^n=c_n\det\left(\frac{\partial^2\psi}{\partial z^\mu\partial\bar{z}^\nu}\right)\text{d}V.
\end{equation}
Therefore, we may write\footnote{Recall that there is a Log$_*$ hidden in the front of the left hand side.}
\begin{equation}
	c_n\det\left(\frac{\partial^2\psi}{\partial z^\mu\partial\bar{z}^\nu}\right)=\frac{\mu(x)}{\nu(\nabla u(x))},
\end{equation}
and we could relate the K\"ahler potential on the geometry side to the Brenier potential on the NN side.

Here, we identify certain manifolds in K\"ahler geometry with certain models in NNs by viewing both of them from the perspective of optimal transport. When defining the dual Hessian manifold, one needs to introduce the so-called affine $\mathbb{R}$-bundle over $\mathcal{M}$. We have not provided anything on the NN side that corresponds to this bundle structure. In \cite{DBLP:journals/corr/abs-1903-03511}, an NN learning system is regarded as a fibre bundle. It might be possible that we could find some correspondence there.

\section*{Acknowledgements}
The authors wish to thank their numerous collaborators on these projects. JB is supported by a CSC scholarship. YHH would like to thank STFC for grant ST/J00037X/2. E Heyes would like to thank SMCSE at City, University of London for the PhD studentship, as well as the Jersey Government for a postgraduate grant. E Hirst would like to thank STFC for the PhD studentship.

\linespread{0.8}\selectfont
\addcontentsline{toc}{section}{References}
\bibliographystyle{utphys}
\bibliography{references}

\end{document}